\documentclass[useAMS,usenatbib,times,letter,amssymb]{mn2e}
\usepackage{epsfig,times,amssymb}

\title[Dark Matter in A901/902]{The dark matter environment of the Abell 901/902 supercluster: a weak lensing analysis of the {\it HST} STAGES survey}
\author[Heymans et al.]{Catherine Heymans$^{1,2}$\thanks{heymans@physics.ubc.ca}, Meghan E. Gray$^3$, Chien Y. Peng$^{4,5}$, Ludovic Van Waerbeke$^{1}$, \newauthor 
Eric F. Bell$^{6}$, Christian Wolf$^{7}$, David Bacon$^8$, Michael Balogh$^9$, Fabio D. Barazza$^{10}$,\newauthor Marco Barden$^{11}$, 
Asmus B\"ohm$^{12}$, John A.R. Caldwell$^{13}$,
Boris H\"au\ss ler$^3$,  
Knud Jahnke$^{6}$, \newauthor  
Shardha Jogee$^{14}$, Eelco van Kampen$^{11}$,  
Kyle Lane$^{3}$, 
Daniel H. McIntosh$^{15}$,\newauthor
Klaus Meisenheimer$^{6}$, Yannick Mellier$^{2}$, 
Sebastian F. S\'anchez$^{16}$,
Andy N. Taylor$^{17}$, \newauthor 
Lutz Wisotzki$^{12}$  \& 
Xianzhong Zheng$^{18}$.\\ 	
$^1$Department of Physics and Astronomy, University of British Columbia, 6224
Agricultural Road, Vancouver, V6T 1Z1, Canada.\\
$^2$Institut d'Astrophysique de Paris, UMR7095 CNRS, 98 bis bd Arago, 75014 Paris, France.\\
$^3$School of Physics and Astronomy, The University of Nottingham, University Park, Nottingham NG7 2RD, UK.\\
$^4$NRC Herzberg Institute of Astrophysics, 5071 West Saanich Road, Victoria, V9E 2E7, Canada.\\
$^5$Space Telescope Science Institute, 3700 San Martin Drive, Baltimore, MD
    21218, USA.\\
$^{6}$Max-Planck-Institut f\"{u}r Astronomie, K\"{o}nigstuhl 17, D-69117,
Heidelberg, Germany.\\
$^{7}$Department of Astrophysics, Denys Wilkinson Building, University of
    Oxford, Keble Road, Oxford, OX1 3RH, UK.\\
$^8$Institute of Cosmology and Gravitation, University of Portsmouth, Hampshire Terrace, Portsmouth, PO1 2EG, UK. \\
$^9$Department of Physics and Astronomy, University Of Waterloo, Waterloo, Ontario, N2L 3G1, Canada.\\ 
$^{10}$Laboratoire d'Astrophysique, \'Ecole Polytechnique F\'ed\'erale de 
Lausanne (EPFL), Observatoire, CH-1290 Sauverny, Switzerland.\\
$^{11}$Institute for Astro- and Particle Physics, University of  
Innsbruck, Technikerstr. 25/8, A-6020 Innsbruck,
   Austria. \\
$^{12}$Astrophysikalisches Insitut Potsdam, An der Sternwarte 16, D-14482 Potsdam, Germany.\\
$^{13}$University of Texas, McDonald Observatory, Fort Davis, TX 79734, USA.\\
$^{14}$Department of Astronomy, University of Texas at Austin, 1 University
    Station, C1400 Austin, TX 78712-0259, USA.\\
$^{15}$Department of Astronomy, University of Massachusetts, 710 North Pleasant
    Street, Amherst, MA 01003, USA.\\
$^{16}$Centro Hispano Aleman de Calar Alto, C/Jesus Durban Remon 2-2, E-04004
    Almeria, Spain.\\
$^{17}$The Scottish Universities Physics Alliance (SUPA), Institute for Astronomy, University of Edinburgh, Blackford Hill, Edinburgh, EH9 3HJ, UK.\\
$^{18}$Purple Mountain Observatory, National Astronomical Observatories, Chinese Academy of Sciences, Nanjing 210008, PR China. \\ 
}

\newcommand{\be}{\begin{equation}}  \newcommand{\ee}{\end{equation}}
\newcommand{\bes}{\begin{equation*}}  \newcommand{\ees}{\end{equation*}}
  \newcommand{\ba}{\begin{eqnarray}}
\newcommand{\ea}{\end{eqnarray}}

\newcommand{\rag}{\rangle}
\newcommand{\lag}{\langle}
\def\gs{\mathrel{\raise1.16pt\hbox{$>$}\kern-7.0pt %
\lower3.06pt\hbox{{$\scriptstyle \sim$}}}}         %
\def\ls{\mathrel{\raise1.16pt\hbox{$<$}\kern-7.0pt %
\lower3.06pt\hbox{{$\scriptstyle \sim$}}}}         %

\begin{document}

\date{Accepted 2008 January 4. Received 2008 January 3; in original form 2007 October 16.}

\pagerange{\pageref{firstpage}--\pageref{lastpage}} \pubyear{2008}
\maketitle
\label{firstpage}

\begin{abstract}
We present a high resolution dark matter reconstruction of the $z=0.165$ Abell 
901/902 supercluster from a weak lensing analysis of the {\it HST} STAGES survey.  
We detect the four main structures of the supercluster at high significance, 
resolving substructure within and between the clusters.  We find that the 
distribution of dark matter is well traced by the cluster galaxies, with the 
brightest cluster galaxies marking out the
strongest peaks in the dark matter distribution.  
We also find a significant extension
of the dark matter distribution of Abell 901a in the direction of an 
infalling X-ray group Abell $901\alpha$.  We present
mass, mass-to-light and mass-to-stellar mass ratio measurements of the 
structures and substructures 
that we detect.  We find no evidence for variation of the mass-to-light 
and mass-to-stellar mass ratio between the different clusters.  We 
compare our space-based lensing analysis with an earlier 
ground-based lensing analysis of the supercluster
to demonstrate the importance of space-based 
imaging for future weak lensing dark matter `observations'. 
\end{abstract}

\renewcommand{\arraystretch}{1.3}

\section{Introduction}

Observations and theory both point to the importance of environment on 
the properties of galaxies. Early-type galaxies are typically
found in more dense regions compared to late-type galaxies \citep{Dressler}, 
galaxy colour and luminosity are found to be closely related to galaxy density 
\citep{Blanton05} and the fraction of
star-forming galaxies also shows a strong sensitivity to the density
on small $<1$ Mpc scales \citep{Balogh04,Blanton06}.  
Theoretically there are a number of physical mechanisms that could cause 
these effects in dense environments.  These processes
can change the star formation history, gas content and/or 
morphology of a galaxy 
through, for example, 
ram-pressure stripping \citep{GG72,Larson,Balogh00} 
and/or the tidal effects of 
nearby galaxies \citep[galaxy harassment,][]{Moore96} and/or the tidal 
effects of the dark matter potential \citep{Bekki,Moore98}.  
They depend differently on 
cluster gas, galaxy density and the dark matter potential however, with
ram-pressure stripping dependent on the gas distribution compared to 
tidal effects which are dependent on the overall potential.  
A key difficulty in disentangling these effects observationally, 
is that typically the tidal potential is only constrained
in a global sense through the measured velocity dispersion of a cluster, or a 
richness or total luminosity estimate.  This results in an assumed 
spherical tidal 
potential model that is smoothed over the small scales that are relevant 
for tidal stripping and harassment studies.

In this paper we study the complex 
Abell 901/902 supercluster, hereafter A901/2, 
in the first of a series of papers from the 
STAGES\footnote{`Space Telescope A901/902 Galaxy Evolution Survey'
({\it HST} GO-10395, PI M. E. Gray), www.nottingham.ac.uk/$\sim$ppzmeg/stages } 
collaboration.
From a rich multi-wavelength dataset the A901/2 supercluster 
permits a thorough
investigation of the relationships between galaxy morphology \citep[from
Hubble Space Telescope ({\it HST}) and ground-based imaging,][]
{GrayPeng,Lane}, luminosity, stellar mass and colour 
\citep[from the COMBO-17 survey with 17-band
optical imaging,][]{Wolf03, Borch}, star
formation rates \citep[from 24$\mu$m {\it Spitzer} data,][]{Bell07},
galaxy density \citep{wgm,Gray04}, 
and the hot
intra-cluster medium \citep[from XMM observations,][]{Gilmour,Gray07}.  
One of the key reasons to obtain {\it HST} imaging of this supercluster was to
construct a high resolution, reliable and accurate
map of the projected total mass density distribution.
Using weak gravitational lensing techniques we are able to reconstruct
the distribution of both dark and luminous matter
and quantify the significance of the structures that are seen, updating
the previous ground-based weak lensing analysis of \citet{meg02}.
This extra dimension to the 
multi-wavelength view of A901/2 will be a key
ingredient in future studies where we hope to be able to 
separate the effects of
tidal and gas-dynamical influence on galaxy formation and evolution.

Weak gravitational lensing is now a well established method for
studying the distribution of dark matter.  Light from distant galaxies
is deflected by the gravitational effect of the intervening structures,
inducing a weakly coherent distortion in the shapes of galaxy images.
The strength of this lensing effect is directly related to the 
projected mass along the line of sight, and it can therefore 
be used to map dark matter in dense regions
\citep[see for example][]{meg02,Gavazzi04,Dietrich,Clowe06,Mahdavi}.

The first weak lensing analysis of A901/2 by \citet{meg02} used
deep ground-based $R$-band observations from the COMBO-17 
survey \citep{Wolf03}.
This analysis
revealed three significant peaks in the dark matter distribution at the 
locations of the A901a, A901b and A902 clusters, in addition to a 
low significance south west
peak co-incident with a galaxy group, hereafter referred to as the SW group. 
This analysis also showed a 
filamentary extension between the A901a and A901b clusters.  As this filament
was located across the CCD chip boundary in the mosaic image, however, 
\citet{meg02} could not rule out the possibility of this structure 
originating from residual uncorrected distortions from the point 
spread function (PSF) of the telescope and detector.
The \citet{meg02} ground-based analysis also 
reported a candidate giant arc. 
The STAGES {\it HST} imaging can rule out this candidate arc as 
a co-incidental alignment of objects.  STAGES does however 
resolve several other candidate arcs around supercluster galaxies, 
which will be presented in \citet{Aragon} and \citet{GrayPeng}.

Using the accurate photometric redshift information from the A901/2 
17-band observations of the COMBO-17 survey, 
where the photometric redshift error 
$\sigma_z \sim 0.02/(1+z)$ for $R<24$, 
\citet{A9013D} extended 
the \citet{meg02} analysis, by creating 
a three-dimensional reconstruction of the A901/2 dark matter 
distribution.  This analysis 
revealed a previously unknown higher redshift cluster located 
behind A902 that is at $z=0.46$.  This cluster was 
named, and hereafter referred to as, CB1 by \citet{A9013D}.   
We have updated 
the redshifts of both CB1 and A901/2 
in this analysis based on an improved photometric redshift catalogue and 
the addition of some spectroscopic redshifts.

In this analysis we revisit the dark matter distribution in A901/2 using 
deep {\it HST} observations.  The dominant source of 
noise in the weak lensing analysis of clusters is the Gaussian noise 
introduced 
from the random intrinsic ellipticities of galaxies.  Weak lensing maps 
of dark matter on small scales therefore benefit greatly from the high 
resolution that {\it HST} has to offer.  {\it HST} triples the number density
of resolved galaxies from which the lensing signal can be measured, 
reducing the intrinsic ellipticity noise on small scales. 
In addition, the high 
resolution space-based data permits higher signal-to-noise shape 
measurements and a narrower PSF, 
thus implying a more accurate PSF correction.  

This paper is organised as follows.  In section~\ref{sec:theory} we describe 
the weak lensing theory that is related to this analysis, and the 
maximum likelihood method that we use 
to reconstruct the dark matter distribution.
We describe the data and weak lensing measurement method in 
section~\ref{sec:data}.  We present our results in 
section~\ref{sec:res}, including NFW profile mass measurements 
in section~\ref{sec:nfwmass} and the dark matter reconstruction
and ground-based comparison
in section~\ref{sec:maps}.  A first comparison of the dark matter,
galaxy light and stellar mass distribution is presented in section~\ref{sec:masslight}
along with mass, mass-to-light  and mass-to-stellar mass ratio measurements.
A more detailed comparison of the mass, gas and galaxies of A901/2
will appear in a forthcoming analysis.
We investigate the significance of the supercluster substructure that is 
resolved in our dark matter reconstruction in section~\ref{sec:peaksig} and
discuss our findings and conclude in section~\ref{sec:conc}.  Throughout 
this paper we assume a $\Lambda$CDM cosmology with $\Omega_m =0.3$, 
$\Omega_\lambda =0.7$, and ${\rm H}_0 = 100 {\rm h}\, {\rm km \, s^{-1} Mpc^{-1}}$.  All magnitudes are given in the Vega system.

\section{Method and Theory}
\label{sec:theory}
Gravitational lensing is sensitive to the projected surface mass
density along the line of sight $\Sigma(\theta)$, 
typically denoted by the convergence
$\kappa$.  In the case of a single lens,
\ba 
\kappa = \frac{\Sigma }{\Sigma_{\rm crit}} \,, &
\displaystyle{\Sigma_{\rm crit} = \frac{{\rm c}^2}{4\pi {\rm G}} \frac{D_s}{D_l D_{ls}}}\, ,
\label{eqn:sc}
\ea
where $D_l$ is the angular diameter distance to the lens, 
$D_s$ is the angular diameter distance to the lensed
source galaxies, 
and $D_{ls}$ is the angular diameter distance from the lens to the source.

The coherent distortion, or reduced shear $g = g_1 +i g_2$, that is
detected in the images of distant sources allows for the
reconstruction of the projected intervening matter $\kappa$ as $g =
\gamma / (1-\kappa)$, and 
\ba 
\kappa = \frac{1}{2} (\psi_{,11} +\psi_{,22})\, , & 
\gamma_1 = \frac{1}{2} (\psi_{,11} - \psi_{,22})\, ,&
\gamma_2 =\psi_{,12} \, , 
\label{eqn:kg}
\ea 
where $\gamma$ is the true shear,
$\gamma = \gamma_1 + i\gamma_2$, and $\psi_{,ij}$ is the second
derivative of the lensing potential \citep[see for example][]{Bible}.

The strength of all lensing distortions is invariant under the
transformation $\kappa^\prime = (1-\lambda)\kappa + \lambda$, where
$\lambda$ is a constant \citep{masssheet}.  
This is known as the `mass sheet degeneracy'
implying that all lensing observations are insensitive to a constant mass
sheet across the field of view ($\lambda$) 
in addition to a $\kappa$ dependent rescaling of the `original' 
surface mass density. For wide-field images of relatively
isolated clusters, where $\kappa$ is weak,
one can significantly reduce this bias using the $\zeta_c$
statistic of \citet{Clowe98}.  The $\zeta_c(r)$ statistic gives a model free
estimate of the mass enclosed within a radius $r$ and is given by
\be 
\zeta_c(r_1) = \bar{\kappa}(r \leq r_1) - \bar{\kappa}(r_2 \leq r \leq r_{\rm max}) \, ,
\label{eqn:zetac}
\ee
where $r_2$ is defined to be the radius outside which the cluster density is 
expected to be very low, based on initial mass estimates, 
and $r_{\rm max}$ is the field-of-view radius.  
The second term therefore essentially measures the constant $\lambda$.  
In the case of A901/2 we find $\bar{\kappa}(15'
\leq r \leq 20') = -0.002 \pm 0.007$
where $r$ is measured from the centre of the
STAGES mosaic which is centred on the supercluster.
This measure is consistent with what would 
be expected from large-scale structure and the NFW multi-halo model of the 
A901/2 supercluster that we develop in section~\ref{sec:nfwmass}.
We therefore assume a zero mass sheet degeneracy 
correction in the analysis that follows.

\subsection{Dark Matter reconstruction}

In this paper we use a maximum likelihood method to reconstruct the
surface mass density $\kappa$.  Starting with a `best guess'
\citet{KS93} reconstruction, the lensing potential $\psi$ is
constructed on a pixelised grid and is allowed to vary to produce the
minimum difference between the reconstructed and observed reduced
shear field.  The benefit of using this method is that a varying noise
estimate can be obtained across the whole region enabling the
significance of each structure in the dark matter map to be accurately
quantified.  Furthermore it does not rely on the assumption that the
observed reduced shear $g$ is approximately equal to the true shear
$\gamma$, which for the A901/2 supercluster would introduce errors at
the $\sim$15\% level.   We smooth the resulting $\kappa$ maps with a Gaussian 
of smoothing scale 0.75 arcmin, which 
is equal to $\sim 90 {\rm h}^{-1}$ kpc at the supercluster redshift $z=0.165$.  
This smoothing 
scale provides the best trade-off between high resolution and high 
signal-to-noise.  

We determine the location of peaks from the local maxima and minima in 
the signal-to-noise weak lensing map.  Occasionally we find two peaks that are 
separated by less than half the smoothing radius.  These arise from small 
noise fluctuations on top of a larger fluctuations and in these cases we 
only count a single peak with significance given by the maximal 
peak within the smoothing radius.
Once peaks are detected in a 
weak lensing mass map their significance has to be compared
to what is expected from a smoothed random noise map, 
where a $3\sigma$ noise peak, for example, is much more common
than would naively be expected.
As shown by \citet{VW2000}, the statistics of peaks in a smoothed 
pure noise map follow the peak statistics of a two-dimensional 
Gaussian random field \citep{BE87}.  We use both the 
peak signal-to-noise and the radial peak profile to calculate the 
global probability of a detected dark matter peak arising from noise 
using Equation (45) of \citet{VW2000}.  

\subsection{Model-free Mass measurement}
As our dark matter reconstruction reveals structures 
that are far from the spherically symmetric simple isothermal sphere and 
NFW models \citep{NFW97} that are often fit to estimate masses from weak 
lensing measurements \citep[see for example][]{HH07},
our preferred method to measure mass 
uses a model-free mass estimate.  Following the idea of the 
$\zeta_c$ statistic (Equation~\ref{eqn:zetac}), 
we measure the mass of structures within an aperture.
For the main structures in the supercluster we define apertures 
by the $1\sigma $ and $3\sigma$ enclosed regions in the 
dark matter signal-to-noise maps.  In the cases of smaller cluster 
substructure, where the smoothed structures appear to be 
more spherical, we use 
circular apertures of radius 0.75 arcmin to match the smoothing scale used 
in the dark matter reconstruction. The `aperture' mass is given by
\be
M = \sum_{\rm aperture} A_{\rm pix}\,\, \kappa(x,y) \,\,\Sigma_{\rm crit}\, , 
\label{eqn:mass}
\ee
where $A_{\rm pix}$ is the projected pixel area at the cluster redshift in 
${\rm h}^{-2} {\rm Mpc}^2$, $(x,y)$ are pixels enclosed by the chosen aperture 
and $\Sigma_{\rm crit}$ is the critical surface mass 
density, given in Equation~\ref{eqn:sc}. 

\subsection{NFW profile model}
\label{sec:nfwtheory}

The main drawback of using the model-free mass estimate in
Equation~\ref{eqn:mass} is the inability 
to separate mass at different redshifts.  This is because
the dark matter reconstruction $\kappa$ measures the 
projected surface mass density along the line of sight.  
In the case of the A901/2 
supercluster there is a higher redshift $z=0.46$ cluster, CB1, that lies
behind A902 \citep{A9013D} such that the model free mass estimate for the A902 
region gives the combined mass of A902 and CB1.  
To obtain separate mass estimates for the A902 and CB1 cluster 
and to enable a comparison to future analyses of numerical simulations, 
we therefore also present mass estimates for the dark matter structures 
in A901/2 using an NFW halo model.  

The NFW halo model has been shown 
in numerical simulations to provide a good fit to the spherically averaged 
profile of all dark matter halos irrespective of their mass \citep{NFW97}.
The NFW model for the density profile of a halo at redshift $z$ is
given by
\be
\rho(r) = \frac{\delta_c \, \rho_c(z)}{(r/r_s)(1 + r/r_s)^2}\,,
\ee
where $\delta_c$ is the characteristic density, $r_s$ is the scale radius and
$\rho_c(z)$ is the critical density given by $3{\rm H}(z) 
^2/8\pi {\rm G}$.
We follow \citet{Dolag04} defining 
the virial radius $r_{200}$ as the radius where the  
mass density of the halo is equal to $200 \Omega_m(z) \rho_c(z)$,  
such that the corresponding virial
mass $M_{200}$ is given by
\be
M_{200} = 200 \, \Omega_m(z) \rho_c(z)\, \frac{4\pi}{3} r_{200}^3 \,.
\label{eqn:Mtriangle}
\ee
As the mass enclosed within a radius 
$R$ is given by 
\be
M(r \leq R) = 4\pi \delta_c\, \rho_c(z) \, r_s^3 \left[ \ln \left( 1 + \frac{R}{r_s} \right) - \frac{R/r_s}{1+R/r_s} \right]\,,
\label{eqn:Mr}
\ee
defining the concentration parameter as $c = r_{200} / r_s$, the
characteristic halo density $\delta_c$ is given by
\be
\delta_c = \frac{200\, \Omega_m(z)}{3} \frac{c^3}{\ln(1+c) - c/(1+c)} \,.
\ee
For a given CDM cosmology, the halo mass $M_{200}$
and concentration $c$ are related \citep{NFW97,Bullock01,Eke01,Dolag04}, 
where the
dependence is calculated through fits to numerical simulations.  
In this paper we use the relationship between halo mass $M_{200}$ 
and concentration $c$ derived by \citet{Dolag04}.

The expression for the weak lensing
shear $\gamma$ and convergence $\kappa$ 
induced by an NFW dark matter halo, given in \citet{Bartelmann96} and 
\citet{WB00}, depends on 
the redshift of both the lens and source galaxies.  In this analysis we have 
accurate redshifts for the majority of the A901/902 cluster galaxies 
but no redshift information 
for $\sim 90\%$ of our source galaxies as they are too faint to calculate a
COMBO-17 photometric redshift.
The maximum-likelihood method of \citet{SchRix} was designed to take  
advantage
of such a data set for analysing the galaxy-galaxy lensing
statistically \citep{MK06,GEMSgg}, 
and it is this method that we have adapted for cluster
lensing and describe below.

For a model cluster density profile, in the case where all galaxy
redshifts are known,
the weak shear $\gamma$ and convergence $\kappa$ experienced by
each source galaxy can be predicted by
summing up the shear and convergence contributions from all
the foreground clusters.
In this analysis the redshifts of the source galaxies are unknown, and
we therefore assign those galaxies a magnitude-dependent redshift
probability distribution $p(z,{\rm mag})$ given by Equation 15 of 
\citet{HymzGEMS} 
updated with the magnitude-redshift relation of \citet{SchPSF}, where
the average median redshift $z_m$ is given by
\be
z_m = 0.29 [m_{\rm F606W}-22] + 0.31\, .
\label{eqn:Sch}
\ee
We are then able to calculate the
expectation value of the observed reduced shear $\lag g \rag$ through Monte
Carlo integration by drawing a source galaxy redshift estimate $z_s^\nu$
from the distribution $p(z,{\rm mag})$, $\nu = 1..N_{\rm MC}$ times,  
where $N_{\rm MC} = 100$ in this analysis.  Testing larger values for 
$N_{\rm MC}$ did not change the result.  For each
$z_s^\nu$ estimate the induced cluster lensing shear $g^\nu$
is calculated with the resulting mean reduced shear given by
\be
\lag g \rag = \frac{1}{N_{\rm MC}} \displaystyle\sum_{\nu=1}^{N_ 
{\rm MC}}g^\nu \,.
\ee
The intrinsic
source galaxy ellipticity $\epsilon^{\rm s}$ is then calculated,
$\epsilon^{\rm s} \approx
\epsilon^{\rm obs} - g$. The distribution of each component of the
observed galaxy ellipticity
is well described, for the STAGES survey,
by a Gaussian of width $\sigma_\epsilon = 0.26$.
As the induced reduced shear $g$ is relatively weak, 
the probability for observing an
intrinsic ellipticity of $\epsilon^{\rm s}$ is then given by
\be
P(\epsilon^{\rm s}) = \frac{1}{2\pi\sigma_\epsilon^2} \exp \left[
   \frac{-|\epsilon^{\rm s}|^2}{2 \sigma_\epsilon^2} \right] \,\, .
\ee
The best-fit dark matter halo parameters are determined
by maximising the likelihood $L = \Pi \left[P(\epsilon^{\rm s})_i 
\right]$
where the product extends over all source galaxies $i$.

\section{The STAGES Data}
\label{sec:data}
The STAGES survey \citep{GrayPeng} spans a quarter square degree 
centered on the A901/2 supercluster.  Imaged in F606W, using the 
{\it HST} Advanced Camera for Surveys (ACS),
the 80 orbit mosaic of 80 ACS tiles forms the second largest 
deep image taken by {\it HST}.  A detailed account of the STAGES 
reduction method 
will be presented in \citet{GrayPeng}.  It is very similar to the reduction 
used for the GEMS survey discussed in \citet{HymzGEMS} and 
\citet{CaldwellGEMS}, differing only in the dither and drizzle strategy.  
For STAGES, each image consists of four co-added 
dithered images combined with a Gaussian drizzling kernel 
with a resulting 0.03 arcsecond pixel scale, as suggested by \citet{RhodesPSF}.
STAGES is complemented by 17-band optical imaging from the COMBO-17 survey 
which provides, for galaxies brighter than $R=24$, 
accurate photometric redshifts with errors 
$\sigma_z \sim 0.02 (1+z)$, spectral energy distribution galaxy 
classification, and stellar mass estimates $M_*$ from low resolution 
17-band spectra fits to parameterised
star formation history models \citep{Borch, C17}.  

\subsection{Weak lensing shear measurement}
\label{sec:WLmeas}

\begin{figure}
\begin{center}
\epsfig{file=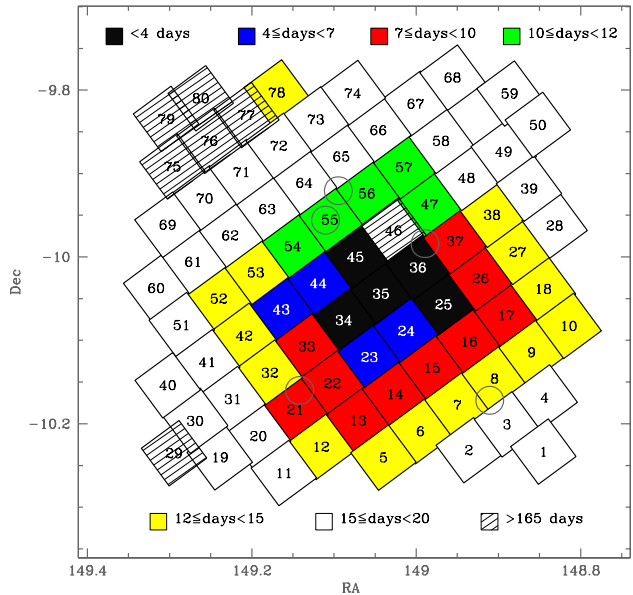,width=8.75cm,angle=0,clip=}
\caption{The tiling pattern of the 
STAGES ACS observations.  Each group of data used to make the seven 
time-dependent PSF models is shown with a different grey-scale. 
The positions of the four main structures are shown with a circle of radius 1 
arcmin centred on the brightest cluster galaxy in A901a (tile 55), 
A901b (tile 36), A902 (tile 21) and the SW group (tile 8).  The position of the
infalling X-ray group A901$\alpha$ is also circled (above tile 55).}
\label{fig:obstile}
\end{center}
\end{figure}

To measure the reduced weak lensing shear $g$, 
we use the data reduction steps and
method described in \citet{RixGEMS} and \citet{HymzGEMS}.  
The shear measurement aspect 
is based on the \citet{KSB} method.
As we are primarily interested in the variation in the dark matter map 
we have updated our shear 
measurement pipeline to maximise the signal-to-noise by including  
a polynomial fit to the shear seeing correction $P^\gamma$ 
\citep{LK97} as a 
function of galaxy size.  We also include the Hoekstra 
correction to the shear polarisability tensor detailed in \citet{HymzSTEP}. 
The accuracy of these updates has been verified using the 
publicly available suite of simulations from the Shear TEsting 
Programme\footnote{www.physics.ubc.ca/$\sim$heymans/step.html} 
\citep{HymzSTEP,STEP2}.  The modifications
successfully reduced the noise on the shear measurement, quantified
through the root-mean-square variation 
of the measured ellipticity $\sigma_\epsilon$, 
from $\sigma_\epsilon=0.31$ to $\sigma_\epsilon=0.26$.

We use the same method as \citet{HymzGEMS} to account for the time variation 
of the ACS PSF, namely to divide the data into sets 
imaged in a short period of time and assume that the temporal variation 
during that time is minimal.  The majority of the A901/2 field was observed 
in the space of 20 days, with the remaining 10\% imaged at a later date 
over the space of 4 days.  Owing to the relatively low galactic latitude of 
the A901/2 field and the resulting high stellar density of $30 - 40$ 
useful stellar images per ACS image, we are able to split 
the data into seven groups to achieve good temporal sampling of the PSF 
distortion. 
This number was chosen to balance between
the need to use as many ACS images as possible 
to maximise the signal-to-noise on the average measured stellar
ellipticity as a function of CCD position, 
whilst requiring
as many time bins as possible to minimise the temporal variation of the
PSF pattern.  Figure~\ref{fig:obstile} shows the tiling pattern of the 
STAGES ACS observations denoting each group of data that was
used to make the seven different PSF 
models.  With this semi-time dependent 
model we find and remove temporal variation during the A901/2 
observations.  Averaged across the ACS field-of-view, this temporal variation 
is at the 1\% level on the measured stellar ellipticity.  
As this variation is more than an order of magnitude 
lower than the weak lensing signal from the A901/2 supercluster our 
semi-time dependent PSF model is more than 
sufficient for this analysis.  We might expect to see low-level 
systematics for the ACS images whose observation date is isolated 
at the start or end of a data group, affecting tiles 
21, 33, 36, 43, 44, 46, 47, 57, 69 and 72.  Indeed in the analysis that 
follows we find $>3\sigma$ B-modes, an indication of systematics
\citep{CNPT02}, in tiles 21, 33, 35, 36 and 57.   
These residual systematics will be taken into account by
including a conservative 
systematic error term, based on the B-mode amplitude, in the analysis that 
follows.  Note that
\citet{SchPSF} and \citet{RhodesPSF} present significantly more advanced 
methods to model the temporal variation of the ACS PSF 
designed for the detection of the weaker lensing signal from
large-scale structure which will be investigated further 
in a future analysis.

In the time since the ACS observations of the GEMS survey
used by \citet{HymzGEMS}, 
the charge transfer 
efficiency (CTE) of the ACS has degraded significantly.  During the CCD 
readout, as the CTE degrades over time, the amount of charge left behind 
increases.  This results in image `tails' developing along 
the readout direction, with the most severe effects seen in the furthest
objects from the readout amplifiers.  As the amount of charge left behind in 
each charge transfer is independent of the pixel count, CTE impacts on the 
shapes of fainter objects more significantly than brighter objects, and 
hence this distortion is not taken into account by the PSF correction.  
We follow \citet{RhodesPSF} by using an empirical CTE correction 
$e_1^{\rm CTE}$ 
for the $g_1$ shear component, along the readout direction, where 
$e_1^{\rm CTE} = {\rm A}\, \Delta y/{\rm SN}$.  
$\Delta y$ is the distance from the 
readout amplifier and SN is a signal-to-noise estimate that we define 
as the ratio of the flux and flux error measurements from SExtractor 
\citep{SExt}.  A is a normalisation constant derived to minimise the 
average measured shear $\langle g_1\rangle$, 
where $g_1 = 2(e_1 - e_1^{CTE})/{\rm Tr}(P^\gamma)$, 
$e_1$ is the PSF corrected galaxy ellipticity and $P^\gamma$ is the shear 
polarisabilty tensor from \citet{LK97}.  For the faintest galaxies 
that are furthest from the readout amplifier, and hence the most 
strongly affected, $e_1^{CTE} = 0.02$, but on average $e_1^{CTE} \sim 0.003$
which is more than an order of magnitude 
lower than the weak lensing signal from the A901/2 supercluster.  
We measure the average $\langle g_1 \rangle$ before applying the CTE 
correction to be $0.004$ and after correction  $\langle g_1 \rangle = 0.00001$. Any residual CTE distortions that remain after the correction are 
therefore very weak in comparison to the supercluster lensing signal.  
As the original 
CTE distortion varies across the ACS field of view and hence across 
the STAGES mosaic, any residual CTE distortions would however be included 
in our B-mode analysis and hence the errors in the results that follow.

\subsection{Galaxy selection and redshift estimation}
As we are interested in the dark matter in A901/2 at a redshift of $z=0.165$, 
we select galaxies that are likely to be at higher redshifts and thus 
lensed by the supercluster.
As the majority of our galaxies are too faint to calculate a
COMBO-17 photometric redshift,  the best option
is to use a magnitude selection based on the relationship between 
median redshift $z_m$ and F606W magnitude derived in \citet{SchPSF} and given 
in Equation~\ref{eqn:Sch}.  To ensure 
that the majority of objects have $z_s>z_{\rm A901/2}$, we select 
galaxies with $m_{\rm F606W}> 23$, corresponding to a median redshift $z_m>0.6$.  
We also include selection criteria chosen to optimise the
accuracy and reliability of the weak lensing shear measurement, selecting
galaxies with S/N $> 5$, magnitude $m_{\rm F606W}< 27.5$, and galaxy size 
$r_h>3$ pixels. Our resulting
weak lensing catalogue includes over 60000 objects, 
or roughly 65 galaxies per square arcmin.  
The average galaxy magnitude of 
this sample is $ \langle m_{\rm F606W}\rangle = 25.7$, 
implying a median redshift 
$z_m \simeq 1.4$.  Assuming a redshift distribution given by 
$n(z) \propto z^2 \exp(-z^{1.5})$ \citep{BaughEfst}, we estimate a $\sim 3\%$ 
contamination of our source galaxy catalogue from objects that are 
foreground to the cluster.  The dilution of the signal by foreground 
galaxies is therefore well within the statistical noise of our analysis.

To calculate the model-free mass estimate in Equation~\ref{eqn:mass}
we place the background source galaxy sample
at one redshift $z_s$ taken to be the median redshift $z_m\simeq 1.4$ of 
the sources.   We choose to use the median, as the high redshift tail
of the redshift distribution of faint galaxies 
is poorly constrained observationally 
\citep[see for example][]{JB07} leading to a potentially 
biased measure of the mean 
redshift.  However note that this choice is fairly unimportant as
for $z_s > 1$ and $z_l = 0.165$, the redshift of 
A901/2, the important distance ratio $D_s/D_{ls}$ in Equation~\ref{eqn:sc}
is fairly insensitive to the value of $z_s$.  
For example, a large increase of $z_s$ from $z_s = 1$ to $z_s = 1.5$ 
increases $D_s/D_{ls}$ by only $\sim 6\%$.  Hence for this deep analysis, 
where the majority of sources have redshifts $z_s > 1$, 
placing all lensed galaxies at one redshift is a good approximation.  
 
\begin{figure}
\begin{center}
\epsfig{file=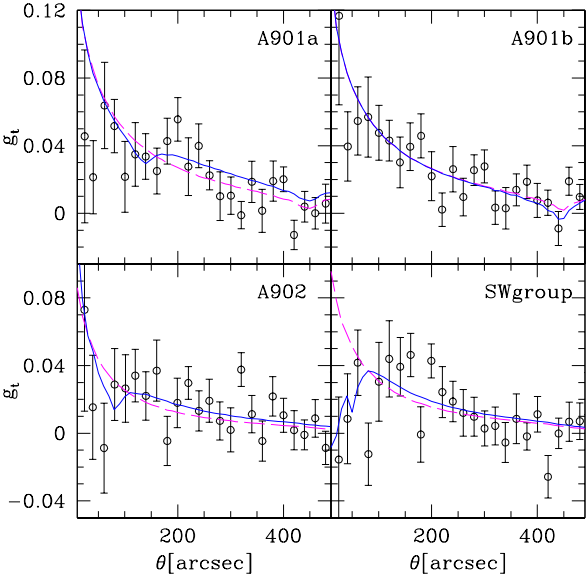,width=8.2cm,angle=0,clip=}
\caption{The tangential reduced shear distortion as a function of the distance from the BCG in each cluster.  The dashed line in each panel shows the best-fitting model profile assuming single NFW dark matter haloes centred on each BCG.  The solid line shows the best-fitting model profile assuming multiple NFW dark matter haloes.  The upper panels show the profile expected from three NFW haloes centred on the BCG in A901a, the BCG in A901b and the X-ray infalling group A901$\alpha$.  A901a and A901$\alpha$ increase the large scale single in A901b and vice versa.  The lower left panel shows the profile expected from two NFW haloes centred on the BCG in A902 and the BCG of the background cluster CB1.  The lower right panel shows the profile expected from two NFW haloes SWa and SWb in the SW group.  The model halo parameters are given in Table~\ref{tab:NFW_res}.}
\label{fig:gt}
\end{center}
\end{figure}

\begin{table*}
\begin{center}
\begin{tabular}{l|l|c|c|c|c|c|c}
\hline
Structure & RA & Dec & $M_{200}$ & $r_{200}$ & $\theta_{200}$ & $M(\theta<1')$ & $\chi^2_{\nu}$\\
& (deg) & (deg)& $({\rm h}^{-1} 10^{13} {\rm M}_{\odot})$ & $({\rm h}^{-1} {\rm kpc})$ & (arcmin) & $({\rm h}^{-1} 10^{13} {\rm M}_{\odot})$\\
\hline
 One Halo: \\
A901a                & 149.1099 &  -9.9561 & $   18.8^{+    4.4}_{-    4.4}$ & $   1194^{+     86}_{-    101}$ & $   10.0^{+    0.7}_{-    0.8}$ & $   1.94^{+   0.19}_{-   0.22}$ & $1.6$\\
A901b                & 148.9889 &  -9.9841 & $   18.1^{+    4.4}_{-    4.4}$ & $   1180^{+     88}_{-    104}$ & $    9.9^{+    0.7}_{-    0.9}$ & $   1.91^{+   0.19}_{-   0.23}$ & $1.0$\\
A902                 & 149.1424 & -10.1666 & $    5.6^{+    2.7}_{-    2.3}$ & $    799^{+    112}_{-    125}$ & $    6.7^{+    0.9}_{-    1.0}$ & $   1.07^{+   0.24}_{-   0.26}$ & $1.3$\\
SW group             & 148.9101 & -10.1719 & $    7.9^{+    3.2}_{-    2.7}$ & $    894^{+    106}_{-    117}$ & $    7.5^{+    0.9}_{-    1.0}$ & $   1.28^{+   0.23}_{-   0.25}$ & $1.9$\\
 \hline Two Halo: \\
A901a                & 149.1099 &  -9.9561 & $   17.5^{+    5.0}_{-    5.0}$ & $   1166^{+    102}_{-    124}$ & $    9.8^{+    0.9}_{-    1.0}$ & $   1.88^{+   0.22}_{-   0.27}$ & $1.8$\\
A901$\alpha$         & 149.0943 &  -9.9208 & $    7.0^{+    3.0}_{-    2.5}$ & $    859^{+    108}_{-    118}$ & $    7.2^{+    0.9}_{-    1.0}$ & $   1.20^{+   0.24}_{-   0.25}$ & -\\
A902                 & 149.1424 & -10.1666 & $    5.0^{+    3.2}_{-    2.3}$ & $    766^{+    137}_{-    140}$ & $    6.4^{+    1.1}_{-    1.2}$ & $   1.00^{+   0.30}_{-   0.29}$ & $1.1$\\
CBI                  & 149.1650 & -10.1728 & $    4.9^{+    3.8}_{-    2.6}$ & $    608^{+    127}_{-    138}$ & $    2.5^{+    0.5}_{-    0.6}$ & $   2.43^{+   1.09}_{-   1.01}$ &-\\
SWa                  & 148.9240 & -10.1616 & $    3.6^{+    2.7}_{-    2.3}$ & $    689^{+    141}_{-    192}$ & $    5.8^{+    1.2}_{-    1.6}$ & $   0.84^{+   0.30}_{-   0.37}$ & $ 1.5$\\
SWb                  & 148.9070 & -10.1637 & $    4.5^{+    2.6}_{-    2.3}$ & $    742^{+    123}_{-    153}$ & $    6.2^{+    1.0}_{-    1.3}$ & $   0.95^{+   0.26}_{-   0.31}$ & -\\

\hline
\end{tabular}
\end{center}
\caption{Mass measurements for the A901/2 supercluster assuming the NFW spherical halo model.  The `one halo' model places a single NFW halo at position (RA,Dec) centred on the BCG in each cluster.  The `two halo' model places a halo at the A901a BCG and the location of the infalling X-ray group A901$\alpha$, a halo at the A902 BCG and at the background cluster CB1 BCG, and two halos in the SW group, SWa and SWb.  There is no motivation to fit the A901b cluster with two haloes 
and it is therefore only listed in the 
`one halo' model upper section of the Table.
The NFW `virial' mass $M_{200} ({\rm h}^{-1} 10^{13} {\rm M}_{\odot})$ corresponds to a `virial' radius $r_{200}({\rm h}^{-1} {\rm kpc})$ which has an observed angular scale $\theta_{200}$ (arcmin).  For comparison with the 1 arcmin aperture model-free mass estimates $M_{\rm ap}$ in Table~\ref{tab:clusmass}, $M(\theta<1')$ is the mass of the NFW halo enclosed by a 1 arcmin aperture, centred on (RA,Dec). The reduced $\chi^2_{\nu}$ of the fit is given in the final column.}
\label{tab:NFW_res}
\end{table*}

\section{Results}
\label{sec:res}
In this section we present the results of our weak lensing analysis of the 
A901/2 supercluster including cluster mass estimates, a comparison of 
three different 
weak lensing dark matter reconstructions and a comparison of the  
resulting dark matter distribution to the 
distribution of light in the supercluster.

\subsection{Mass estimates for NFW profiles}
\label{sec:nfwmass}

We use spherical NFW haloes to model the weak lensing shear measured in the A901/2 field.  We test two different models using the method described in section~\ref{sec:nfwtheory}.  The `one halo' model centres a single NFW halo on the brightest cluster galaxy (BCG\footnote{In this analysis we define a BCG to be the brightest cluster galaxy within an arcminute of the peak of the cluster's galaxy distribution.}) in each cluster at $z=0.165$.  The `two halo' model places a halo at the A901a BCG and the location of the infalling X-ray group A901$\alpha$, a halo at the A902 BCG and at the background cluster CB1 BCG at $z = 0.46$, and two halos in the SW group, SWa and SWb.  The positions of the two SW halos are motivated by the dark matter reconstruction presented in section~\ref{sec:maps}.  Table~\ref{tab:NFW_res} lists the resulting constraints on the NFW `virial' mass $M_{200}$ of each halo, the NFW `virial' radius $r_{200}$ and the corresponding angular projection of this radius on the sky $\theta_{200}$.  We find results that are fully consistent with the single and multiple simple isothermal sphere halo analysis of \citet{A9013D}.  For comparison with the model-free mass estimates in section~\ref{sec:maps} we also calculate the NFW halo mass enclosed by an aperture of 1 arcmin $M(\theta<1')$ using Equation~\ref{eqn:Mr}.  

Figure~\ref{fig:gt} compares the measured tangential reduced shear distortion $g_t$ around each cluster with the prediction from our `one halo' (dashed) and `two halo' (solid) model, assuming all lensed source galaxies are at a single redshift $z_s = 1.4$.  The rotated shear $g_r$ is found to be consistent with zero on all scales as expected.  Note that the model parameters are constrained using the method of \cite{SchRix} as described in section~\ref{sec:nfwtheory}, not as a fit to the azimuthally averaged shear $g_t$ presented in this Figure. We find that both models fit this data equally well but, in the cases of A901a and the SW group, NFW haloes are in general a poor fit to the data (the reduced $\chi^2_{\nu}$ of the fit is given in the final column of Table~\ref{tab:NFW_res}).  From this we conclude that for these unvirialised systems, the spherical NFW model provides a poor fit.  Note that using the \citet{Bullock01} relationship between virial mass and concentration results in even poorer fits to the data.
Figure~\ref{fig:gt} is also instructive to see the impact of the individual halos on each other.  The effect of A901a on the profile of A901b (and vice versa) can be seen from the increasing signal on large scales in the upper panels of Figure~\ref{fig:gt}.   The decrease in signal on small scales in the SW group data favours the `two halo' model over the `one halo' model. 

Comparing the A902 `one halo' and `two halo' models in the lower left panel of Figure~\ref{fig:gt} shows that the addition of the background cluster CB1 has only a weak effect on the profile of A902.   We find that the virial mass of A902 decreases by 11\% when the CB1 halo is included in the analysis (see Table~\ref{tab:NFW_res}).  As CB1 and A902 are separated by 1.4 arcmin, the mass enclosed by a 1 arcmin aperture $M(\theta < 1')$ is even 
less effected and decreases by 6\% when the CB1 halo is included in the analysis.  We therefore conclude from this NFW analysis that in the model-free mass estimates that follow, the contribution from CB1 to the A902 mass cannot be more than a $\sim 10\%$ effect which is within our B-mode systmatic errors in the analysis that follows.

\begin{figure*}
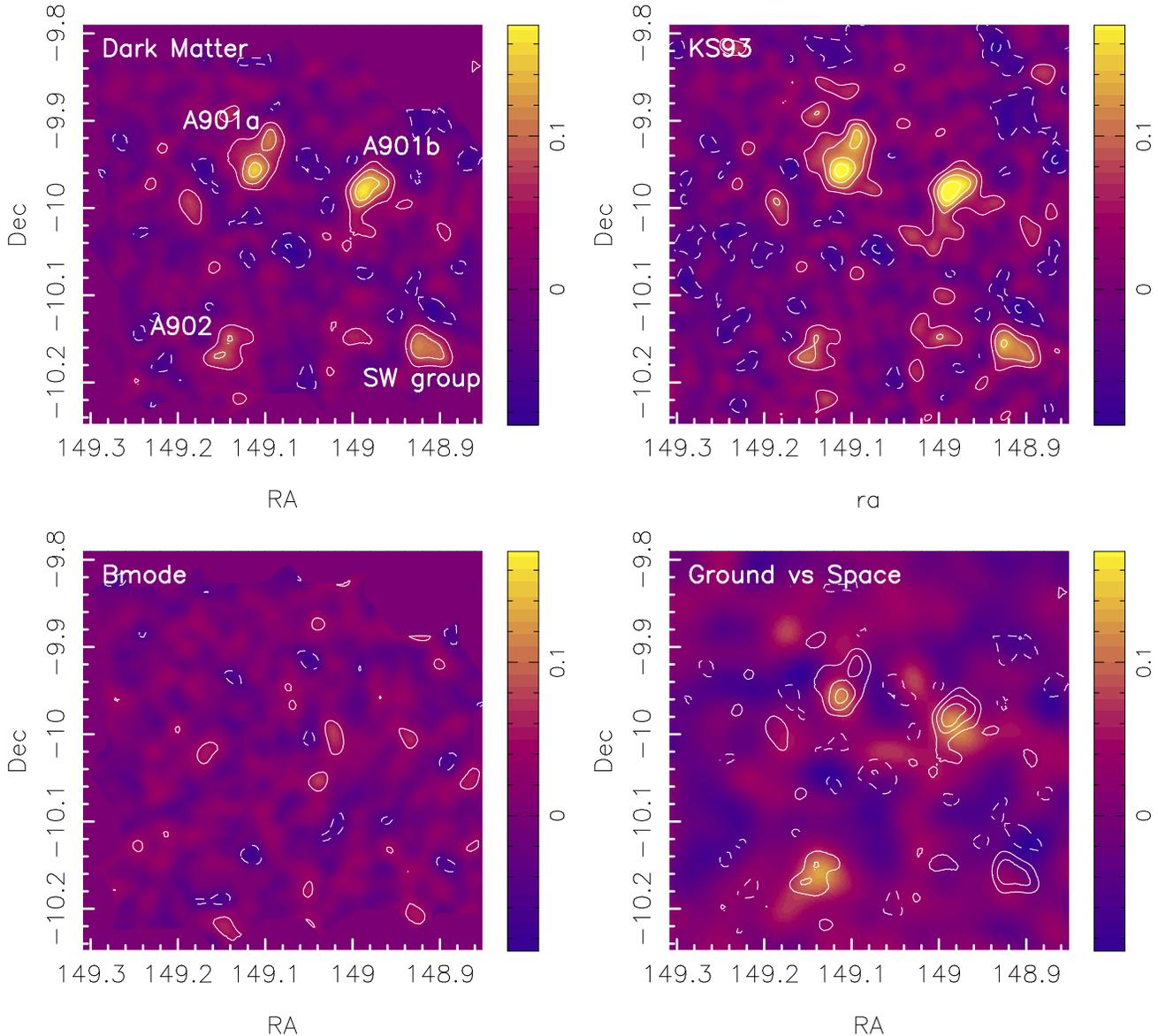

\begin{center}
\begin{tabular}{cc}
\epsfig{file=DM_E_map.ps,width=7.6cm,angle=270,clip=}&
\epsfig{file=DM_KS93_map.ps,width=7.6cm,angle=270,clip=}\\
\epsfig{file=DM_B_map.ps,width=7.6cm,angle=270,clip=}&
\epsfig{file=Ground_Space.ps,width=7.6cm,angle=270,clip=}\\
\end{tabular}
\end{center}
\caption{The dark matter reconstruction of the A901/2 supercluster.  The maximum likelihood dark matter map is shown in the upper left panel clearly revealing the four main superclusters structures; A901a, A901b, A902 and the SW group.  This reconstruction can be compared to a \citet[][KS93]{KS93} reconstruction (upper right panel), the associated B-mode or systematics reconstruction (lower left panel) and the \citet{meg02} ground-based dark matter reconstruction (lower right panel).  The contours enclose the -4$\sigma$,-2$\sigma$ (dashed), 2$\sigma$, 4$\sigma$ and 6$\sigma$ detection regions (solid) of the STAGES maximum likelihood reconstruction (upper right and repeated on the lower left panel), the Bmode reconstruction (lower right panel) and the KS93 reconstruction (upper left panel).  The scale bar shows the amplitude of the mapped convergence $\kappa$ which has been smoothed on 0.75 arcminute scales corresponding to $\sim 90 {\rm h}^{-1}$ kpc at the supercluster redshift $z=0.165$.  The edges of the maximum likelihood dark matter map and the corresponding B-mode (left hand panels) go to zero as a result of the data mask that is shown in the tiling pattern of Figure~\ref{fig:obstile}. }
\label{fig:maps}
\end{figure*}

\subsection{Dark Matter maps}
\label{sec:maps}

Figure~\ref{fig:maps} shows the STAGES dark matter reconstruction of the 
A901/2 supercluster.  The upper left panel shows the maximum 
likelihood reconstruction that clearly reveals 
the four main supercluster structures; A901a, A901b, A902, and the 
SW group.  The contours on this signal-to-noise dark 
matter map correspond to -4$\sigma$,-2$\sigma$ (dashed), 2$\sigma$, 4$\sigma$, 
and 6$\sigma$ detection regions (solid).
Assuming constant noise across the image, the 
scale bar shows the measured convergence $\kappa$.  This is a very
good assumption except for the edges of the map where the noise increases 
rapidly.  

The dark matter map can be compared to the B-mode or `systematics 
map' in the lower left panel of Figure~\ref{fig:maps}.  This is created by 
rotating the galaxies by $45^\circ$ \citep{CNPT02} and reconstructing the 
map.  As weak lensing produces curl-free or E-mode distortions, 
a significant detection of a curl or B-mode signal indicates that 
ellipticity correlations exist from residual systematics.  
Comparing the B-mode map with the contours from the dark matter map therefore 
allows one to assess the reliability of each detected structure.  
For the maps shown in Figure~\ref{fig:maps}, a $3\sigma$ detection has 
$\kappa \approx 0.07$, although the true significance of any peak in the 
distribution has to be determined 
by comparison to the statistics of a random Gaussian field \citep{VW2000}.
For a field this size, with the same number of galaxies, ellipticity 
distribution and smoothing scale, 
smoothed Gaussian noise would produce $2\pm 3$ random $>3\sigma$
E and B-mode peaks, and $0 \pm 1$ random $>3.5\sigma$
E and B-mode peaks,
which we discuss further in section~\ref{sec:peaksig}. 
All but one of the three most significant $>3.5\sigma$ B-mode peaks can be 
linked to regions where the simple semi-time-dependent
PSF modeling used in this analysis fails, as discussed in 
section~\ref{sec:WLmeas}.

For comparison with previous analyses, the upper right 
panel of Figure~\ref{fig:maps} shows a \citet[][KS93]{KS93} reconstruction.  
The main difference seen between our preferred maximum likelihood 
reconstruction (upper left panel) and the KS93 reconstruction 
(upper right panel) is the strength and significance of the peaks.  
In the cluster cores, where $\kappa>0.12$, the  
reduced shear $g$ is more than 15\% larger than the true shear $\gamma$.
Hence the KS93 $\kappa$ reconstruction where 
the reduced shear $g$ is assumed to be
equal to the true shear $\gamma$, results in an overestimation of $\kappa$
in these regions.  When the maps are smoothed, these strong overestimated 
$\kappa$ peaks are smeared out, and in the case of the neighbouring 
A901a and A901b and a large smoothing scale, 
this smearing could lead to false filamentary features that we start to see
between A901a and A901b in the KS93 reconstruction.  
It is this effect, in addition to 
the possible presence of residual PSF systematics, 
that we conclude are responsible for the filamentary 
extension between A901a and A901b seen in the COMBO-17 dark matter 
reconstruction of \citet{meg02}.  

The lower right panel of 
Figure~\ref{fig:maps} compares our STAGES {\it HST} dark matter map (contours), 
with the COMBO-17 ESO 2.2m Wide-Field Imager dark matter map of \citet{meg02}.
This comparison shows excellent agreement in the locations of the dark 
matter peaks.  The ground-based map is 
shown on the same $\kappa$ scale 
as the other maps, but smoothed with a 1 arcmin Gaussian, compared to the 
0.75 arcmin Gaussian used in the STAGES. Increasing the resolution of the 
ground-based map by narrowing the smoothing scale increases the noise 
in the map, thereby further lowering the significance of the detected peaks.
In the ground-based analysis of \citet{meg02}
21 galaxies per square arcmin were used for the dark matter reconstruction
with a root-mean-square variation of the measured ellipticity of $\sigma_\epsilon = 0.44$. 
This can be compared to the 65 galaxies per square arcmin used in this analysis
with $\sigma_\epsilon = 0.26$.  
The increased number density of objects in this analysis results from the high {\it HST} image 
resolution.  The reduction in $\sigma_\epsilon$ results from 
both the higher average signal-to-noise imaging of the source galaxy sample 
and the higher average galaxy-to-PSF size ratio 
that can be achieved with space-based imaging.    
For a $1' \times 1'$ area of sky, the random intrinsic 
ellipticity noise on the measured shear is $\sigma_\gamma(1') = 0.1$ 
for the ground-based analysis and $\sigma_\gamma(1') = 0.03$ for 
this space-based analysis.  As the weak lensing signal that
is typically detected around $z \sim 0.2$ 
clusters is $\gamma \sim 0.1$ and of the order of the ground-based noise,
this comparison clearly 
demonstrates the need for space-based observations for better than
arcminute resolution `imaging' of the dark matter.  

\subsection{A comparison of mass and light}
\label{sec:masslight}

\begin{figure}
\begin{center}
\epsfig{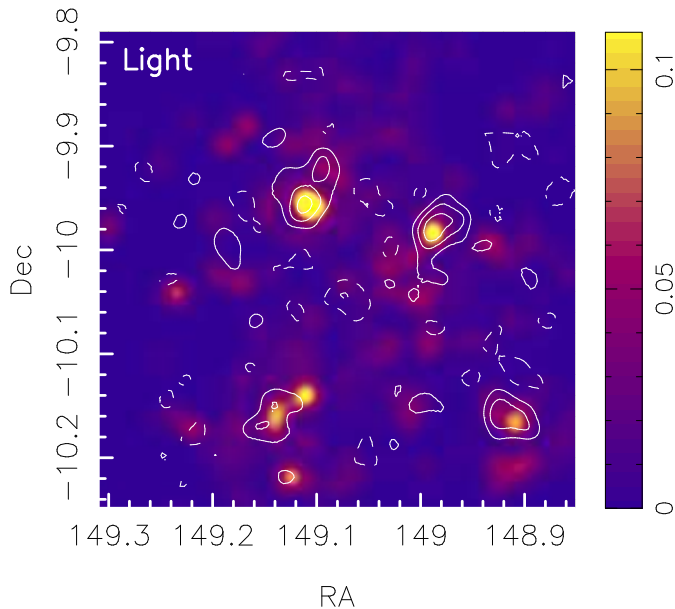}\\
\caption{A comparison of mass and light in A901/2. The mass distribution from Figure~\ref{fig:maps} (shown contoured) is compared to the smoothed light distribution of the cluster galaxies. The scale bar shows the $r$-band luminosity per square arcmin in units of ${\rm h}^{-2} 10^{10}L_{r\odot}$.}
\label{fig:dm_light}
\end{center}
\end{figure}

The distribution of dark matter in A901/2 is found to be very 
well traced by the distribution of galaxies associated with the supercluster, 
as shown by Figure~\ref{fig:dm_light}.  This Figure shows the total $r$-band 
luminosity of cluster galaxies, smoothed on the same scale as the 
dark matter map (shown with contours).
Cluster galaxies are identified from 
the ground-based multi-colour COMBO-17 data using 
the selection criteria from \citet{wgm}; their photometric redshift must 
lie in the range
$0.155 <z_p < 0.185$ and their absolute $V$-band magnitude ${\rm M_V} < -17$, 
which corresponds to an apparent $R$-band magnitude ${\rm R} \lesssim 21.5$.  
These criteria were chosen to keep field contamination low and cluster 
completeness high, with a sample that is 68\% complete at this 
luminosity limit.  
\cite{wgm} estimate the level of field galaxy contamination to be 3\% 
for the red-sequence galaxies, and 15\% for the blue-cloud galaxies, 
\citep[see][for more details]{wgm}.

\begin{figure}
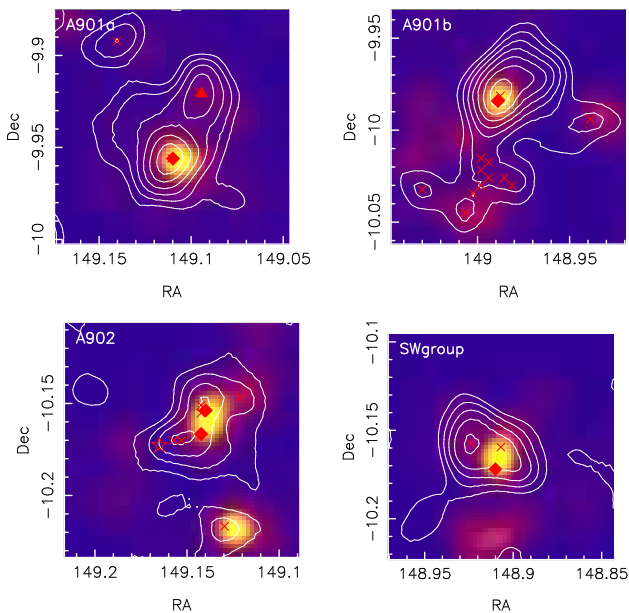

\begin{center}
\begin{tabular}{cc}
\epsfig{file=A901a.ps,width=3.85cm,angle=270,clip=}&
\epsfig{file=A901b.ps,width=3.85cm,angle=270,clip=}\\
\epsfig{file=A902.ps,width=3.85cm,angle=270,clip=}&
\epsfig{file=SWgrp.ps,width=3.85cm,angle=270,clip=}\\
\end{tabular}
\end{center}
\caption{A comparison of mass and light in the main structures of the A901/2 supercluster; A901a (upper left), A901b (upper right), A902 (lower left) and the SW group (lower right).  $1\sigma$ to $7\sigma$ contours of the signal-to-noise dark matter map shown in Figure~\ref{fig:maps} are drawn over a smoothed luminosity map of the old red sequence supercluster galaxies.  The locations of the brightest cluster galaxies are shown (filled diamonds), in addition to the location of the in-falling X-ray group A901$\alpha$ (filled triangle), and the location of the higher redshift $z = 0.46$ cluster CB1 (star). Local maxima in the dark matter map are shown with a cross.}  
\label{fig:mainstruct}
\end{figure}

Immediately in Figure~\ref{fig:dm_light} we can see
that the most massive regions are also the most luminous.
We also start to see cluster 
substructures repeated in both the dark matter and light maps.
A comparison of mass and stellar mass in the cluster is nearly identical to  
Figure~\ref{fig:dm_light}, 
implying that the stellar mass is also a good tracer of 
the underlying dark matter distribution.
Figure~\ref{fig:mainstruct} shows an $8\times 8$ arcmin close-up of the 
four main structures of the A901/2 supercluster.  This Figure compares the 
distribution of dark matter (shown contoured) 
to the luminosity weighted distribution of old 
red sequence galaxies defined in \citet{wgm}.
The locations of the BCGs are shown with 
filled diamonds.
For A901a and A901b, the maximal peak in the dark matter distribution is 
practically co-incident with the location of the BCG, (within 0.25 arcmin).  
For A902 we find two peaks in the dark matter distribution matching the two 
BCGs.  The dark matter peaks are slightly offset from the BCGs (0.5 and 1 
arcmin) due to the presence of CB1, the 
background cluster at a 
redshift of $z=0.46$ whose BCG
location is shown in the A902 
lower left panel of Figure~\ref{fig:mainstruct} with a star.  The cluster 
CB1 fills $\sim 1$ arcmin aperture around the BCG.
The NFW `two halo' A902 and CB1 model detailed in section~\ref{sec:nfwmass}
predicts a shift in the observed 
A902 dark matter peak by $\sim 0.3$ arcmin which is 
consistent with what we find in the dark matter map.

For the SW group,
there is again good agreement with the position of the peak in the mass 
distribution and the BCG, although for this group there are two local 
maxima in the dark matter distribution. Interestingly there are two 
distinct groups in the galaxy population of the SW group.  There is 
an old red galaxy population that surrounds the BCG, as shown in the lower 
right panel of Figure~\ref{fig:mainstruct}.  In addition there is
a dusty red galaxy population 
\citep[described by][but not shown in the Figure]{wgm} that 
exists to the east of the BCG and co-incides with the most massive
eastern dark matter peak 
(denoted SWb in Table~\ref{tab:NFW_res}).  
A more detailed analysis of the interesting 
relationship between the dark matter 
environment and the different galaxy populations will be presented in a 
future paper.  The lower right panel of Figure~\ref{fig:mainstruct} 
also shows one case of a significant 
density of old red galaxies without
a peak in the dark matter distribution.  Towards the edge of the 
STAGES imaging, the noise in our dark matter map grows rapidly, 
and at the location of this galaxy group the noise is twice the noise level
at the SW group.  As this group is likely to be less 
massive than the SW group, which is detected at $5\sigma$, we are not 
surprised that this group is undetected in our dark matter map. 

The A901a upper left panel of Figure~\ref{fig:mainstruct} shows a significant
extension of the dark matter distribution, 
in the direction of the in-falling X-ray group A901$\alpha$ found 
by \citet{Gray07} (shown with a filled triangle).   
The peak along the extension, with $\kappa_{\rm peak} > 4\sigma$ 
(shown with a cross), is co-incident with the brightest galaxy in the 
the X-ray group.

Comparing the local maxima in the A901b 
and A902 distribution with the light maps we find that the 
substructure in the dark matter maps
are often associated with substructures in the 
galaxy distribution.  The only striking discrepancy is a luminous peak to the 
north west of A902, seen in Figure~\ref{fig:dm_light}.  This luminous peak 
results from a single, very luminous dusty red galaxy 
that is brighter than the BCG and is
likely to be infalling on A902 \citep{wgm}.

\begin{figure}
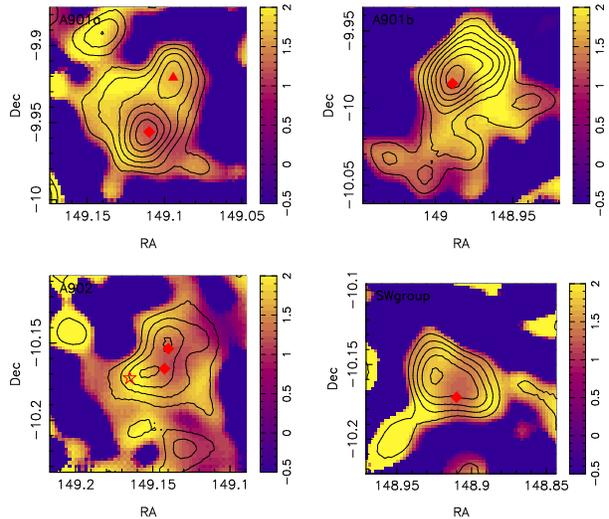

\begin{center}
\begin{tabular}{cc}
\epsfig{file=A901a_MMS.ps,width=3.25cm,angle=270,clip=}&
\epsfig{file=A901b_MMS.ps,width=3.25cm,angle=270,clip=}\\
\epsfig{file=A902_MMS.ps,width=3.25cm,angle=270,clip=}&
\epsfig{file=SWgrp_MMS.ps,width=3.25cm,angle=270,clip=}\\
\end{tabular}
\end{center}
\caption{A comparison of mass and the mass-to-stellar mass ratio $M/M_*$ in the main structures of the A901/2 supercluster; A901a (upper left), A901b (upper right), A902 (lower left) and the SW group (lower right).  $1\sigma$ to $7\sigma$ contours of the signal-to-noise dark matter map shown in Figure~\ref{fig:maps} are drawn over a smoothed mass-to-stellar mass ratio map.  The locations of the brightest cluster galaxies are shown (filled diamonds), in addition to the location of the in-falling X-ray group A901$\alpha$ (filled triangle), and the location of the higher redshift $z = 0.46$ cluster CB1 (star). The scale bar shows $\log(M/M_*)$.}  
\label{fig:MMS}
\end{figure}

In Table~\ref{tab:clusmass} we list mass and mass-to-light ratios for the 
main structures shown in Figure~\ref{fig:mainstruct}.  As discussed in 
section~\ref{sec:nfwmass}, these structures are far from the spherically 
symmetric NFW models that are often used to constrain models.  
We therefore use a model-free mass estimate given by Equation~\ref{eqn:mass}, 
defining the enclosed region using the $1\sigma$ and $3\sigma$ contours 
shown in Figure~\ref{fig:mainstruct}.  For comparison with the ground-based 
analysis of \citet{meg02} and the NFW analysis of section~\ref{sec:nfwmass}
we also list the mass 
enclosed by a 1 arcmin circular aperture (denoted `ap') 
centered on each clusters BCG.  To estimate the contribution of 
systematic error to our mass estimate we follow the conservative
prescription that is 
often used in the analysis of weak lensing by large-scale structure 
\citep[see for example][]{JB07}, calculating errors by 
adding the random error 
(listed as the first mass error in Table~\ref{tab:clusmass})
in quadrature 
with the B-mode signal, shown in the lower left 
panel of Figure~\ref{fig:maps} and listed as the second mass error 
in Table~\ref{tab:clusmass}.  The systematic error dominates the random 
error in this analysis.

\begin{table}
\begin{center}
\begin{tabular}{l|c|c|c|c|c|c}
\hline
Region & $M$ & $M/L$ & $M/M_*$ \\
sig, area & $(\rm{h}^{-1} \, 10^{13}M_\odot)$ 
& $(\rm{h} \, M_\odot / L_{r\odot})$ & $(\rm{h} = 0.7)$\\\hline
A901a   \\
$   1\sigma$,   16.6& $   6.09\, \scriptstyle{\pm   0.07\pm   1.56}$ & $  130.6 \pm   33.4$ & $   31.9 \pm    8.2$ \\
$   3\sigma$,    7.4& $   3.93\, \scriptstyle{\pm   0.04\pm   0.58}$ & $  128.9 \pm   19.0$ & $   29.5 \pm    4.3$ \\
ap, 3.1 & $   1.91\, \scriptstyle{\pm   0.03\pm   0.30}$ & $   93.8 \pm   14.8$ & $   20.6 \pm    3.3$ \\
A901b   \\
$   1\sigma$,   20.4& $   6.52\, \scriptstyle{\pm   0.07\pm   2.16}$ & $  165.2 \pm   54.8$ & $   42.1 \pm   14.0$ \\
$   3\sigma$,    6.2& $   3.59\, \scriptstyle{\pm   0.04\pm   0.57}$ & $  206.1 \pm   33.0$ & $   50.6 \pm    8.1$ \\
ap, 3.1 & $   1.99\, \scriptstyle{\pm   0.14\pm   0.25}$ & $  148.6 \pm   21.7$ & $   34.7 \pm    5.1$ \\
A902    \\
$   1\sigma$,   12.0& $   3.25\, \scriptstyle{\pm   0.05\pm   1.10}$ & $  107.3 \pm   36.5$ & $   27.3 \pm    9.3$ \\
$   3\sigma$,    2.8& $   1.22\, \scriptstyle{\pm   0.03\pm   0.14}$ & $  122.1 \pm   14.4$ & $   28.9 \pm    3.4$ \\
ap, 3.1 & $   1.21\, \scriptstyle{\pm   0.14\pm   0.16}$ & $  107.9 \pm   18.7$ & $   24.4 \pm    4.2$ \\
SWgroup \\
$   1\sigma$,   11.3& $   3.81\, \scriptstyle{\pm   0.05\pm   1.34}$ & $  175.5 \pm   61.5$ & $   40.5 \pm   14.2$ \\
$   3\sigma$,    4.8& $   2.35\, \scriptstyle{\pm   0.03\pm   0.49}$ & $  155.6 \pm   32.3$ & $   34.7 \pm    7.2$ \\
ap, 3.1 & $   1.25\, \scriptstyle{\pm   0.15\pm   0.23}$ & $  126.3 \pm   27.9$ & $   26.1 \pm    5.8$ \\

\hline
\end{tabular}
\end{center}
\caption{Model-free mass measurements for the main structures in the A901/2 supercluster.  The enclosed masses $M$, mass-to-light ratios $M/L$ and mass-to-stellar mass ratios $M/M_*$ are given for the regions defined by the $1\sigma$ and $3\sigma$ contours in the signal-to-noise maps.  The area of these regions is given in the second column (square arcmin).  For comparison with the ground-based analysis of \citet{meg02} we also list the mass and mass-to-light enclosed by a 1 arcmin circular aperture (denoted `ap') centered on the cluster BCG.  The quoted mass errors are listed with the random noise error followed by a conservative estimate of the systematic error. The mass-to-light ratio error and mass-to-stellar mass ratio error include only the error on the mass with the random and systematic parts added in quadrature.}
\label{tab:clusmass}
\end{table}

We find A901a and A901b to be the most massive systems in the supercluster 
with masses $\sim 6.5 \rm{h}^{-1} 10^{13}M_\odot$ and 
mass-to-light ratios $\sim150$ for the full extended $1\sigma$ region.
A902 and the SW group have similar masses, roughly half the 
mass of the A901 pair at $\sim 3.5 \rm{h}^{-1} 10^{13}M_\odot$.  
We find A901b to be the most extended structure in the system, 
and the SW group is the most compact.  

The mass-to-stellar mass ratios
$M/M_*$ of each structure are given in the final column of 
Table~\ref{tab:clusmass}.  These mass ratios were calculated with a Hubble 
parameter ${\rm h}=0.7$, assuming a \citet{Kroupa} initial mass function \citep{Borch}.  
The results are equivalent to within 10\% of the same result
derived using a \citet{Chabrier} or a \citet{Kroupa01} 
initial mass function.  
We find mass-to-stellar mass ratios $M/M_*$ that are 
similar to the ratios found for massive elliptical 
galaxies at this redshift \citep{HHbf05,RM06,GEMSgg}, 
although a direct comparison is 
hard to draw as the results from the massive elliptical galaxies measure 
NFW virial mass to stellar mass ratios instead of the model free mass 
ratio estimates that we present here.  

Figure~\ref{fig:MMS} shows the variation of the mass-to-stellar mass ratio across 
each of the main structures in A901/2 on a log scale, compared to the mass distribution 
(shown contoured). Note that negative regions in the mass reconstruction have been 
set to zero 
in this Figure.  Moving out from the central BCG (shown with diamonds), we find that
the mass-to-stellar mass ratio initially increases, as the stellar mass decreases 
more rapidly than the halo mass.  Continuing out further, 
the mass-to-stellar mass ratio then rapidly decreases as the 
dark matter mass tends to zero.  This Figure shows some regions of 
very high mass-to-stellar mass ratio regions ($\log M/M_* > 2$), but the reader should
note the significance of the mass detected in these regions (shown contoured)
which is less than $2\sigma$ in all cases.

Comparing our convergence $\kappa$ mass reconstruction results to our 
NFW shear analysis, we find very good agreement 
between the mass measured within 
1 arcmin of each clusters BCG which provides an important verification
of our dark matter reconstruction method.
Indeed we find this 
good agreement between the two methods 
continues out to a radius of 4 arcmin, after which 
contribution from to the $\kappa$ map from neighbouring clusters 
complicates the comparison.  
The difference that is seen between the 
NFW virial masses quoted in Table~\ref{tab:NFW_res} 
and the masses quoted in Table~\ref{tab:clusmass} is only a result of the 
different physical 
scales probed in both Tables, which can be seen by comparing the 
observed NFW virial scale $\theta_{200}$ with the region area quoted in the 
first column of Table~\ref{tab:clusmass}.

In comparing our results to the previous ground-based lensing analysis 
of the A901/2 supercluster we must first consider the 
assumption made by \citet{meg02}
that $\kappa \ll 1$ and hence 
$\gamma = g$, where $g$ is the measured reduced shear given above 
Equation~\ref{eqn:kg}.  For the main structures in A901/2  
this would result in an overestimate of cluster mass by $\sim 15\%$.  
Taking this overestimate into account, our mass estimates are consistent 
with \citet{meg02} as can be seen from the space/ground mass map comparison 
in the lower right panel of Figure~\ref{fig:maps}.
The mass-to-light ratio measurements for A901b and A902 disagree 
however at the $3\sigma$ level.
The difference arises from improvements in the selection of the 
cluster galaxies from the COMBO-17 data, in comparison to the previous
two-band optical cluster selection of \citet{meg02}.  
This improvement removes the striking difference between the cluster 
mass-to-light ratio measurements found by \citet{meg02}.  Our results 
show a mass-to-light ratio within an aperture of 1 arcmin
of $M(<1') \sim 100 \rm{h}^{-1} {\rm M}_\odot{\rm L}_\odot $ 
to be a good description for all the main structures in the supercluster.
We find a similar result for the mass-to-stellar mass ratio where  
$M/M_*(<1') \sim 25$ for all the main structures in the supercluster.

\subsection{Supercluster substructure}
\label{sec:peaksig}

In this section we investigate the lower significance peaks 
in the dark matter distribution that are not associated with the cores 
of the supercluster structures discussed above.  Table~\ref{tab:peakstat} 
lists the number of local maxima and minima in the dark matter reconstruction 
for different significance levels and compares them to
what we find in our B-mode reconstruction and what we would 
expect from a smoothed random Gaussian field using Equation (41) from 
\citet{VW2000}.  The high significance peaks $\kappa_{\rm peak} > 4\sigma$ 
are all associated with the cores of the four main supercluster structures.  
However we can see that we have a significant number of  
$\kappa_{\rm peak} > 2\sigma$ peaks that cannot be explained by 
random noise alone.  There are a comparable number of 
$|\kappa_{\rm peak}| > 2\sigma$ peaks in the B-mode map, but comparing the 
location of E and B-mode peaks allows one to assess the reliability of 
the lower significance E-mode detections.   

In order to distinguish between noise peaks and true peaks in the mass 
distribution, it is useful to add morphological information about the 
profile of the peaks.  The mean profile and dispersion of a noise peak 
is given by Equation (47) of \citet{VW2000}.  Comparing the measured 
profile around each detected peak with the mean noise profile allows for 
the calculation of the probability that a peak with a given significance 
and shape is a noise fluctuation, (using Equation (45) of \citet{VW2000}).  
In Table~\ref{tab:peakstat} we list the number of peaks that have a less 
than 33\% probability of being a random noise fluctuation.  The 
result is consistent with the difference between the total number of 
detected peaks and the expected number of random noise peaks.

\begin{table}
\begin{center}
\begin{tabular}{l|c|c|c|c|c|c|c}
\hline
& \multicolumn{3}{c}{$\kappa_{\rm peak} <$} &
\multicolumn{4}{c}{$\kappa_{\rm peak} >$} \\
&  -5$\sigma$& -4$\sigma$& -2$\sigma$& 
       2$\sigma$& 4$\sigma$& 5$\sigma$& 7$\sigma$  \\
\hline
Noise & $   0$ & $   0\pm   0$ & $  19\pm   5$ & $  19\pm   5$ & $   0\pm   0$ & $   0$ & $   0$ \\ \hline
Signal &   0&   1&  36&  29&   7&   3&   1\\
Bmode &   0&   0&  16&  21&   0&   0&   0\\ \hline
Signal (p) &   0&   1&  19&  25&   7&   3&   1\\
Bmode (p) &   0&   0&  10&  14&   0&   0&   0\\ \hline
Signal (b) &   0&   1&  19&  23&   7&   3&   1\\

\hline
\end{tabular}
\end{center}
\caption{Peak statistics; comparing the number of peaks as a function of significance in a smoothed random Gaussian field (Noise), the reconstructed dark matter map (Signal) and the reconstructed B-mode map.  We use the peak profile to further distinguish between a noise peak and a true peak.  Signal (p) and B-mode (p)  list the number of peaks in our substructure sample,
where the peak has less than 33\% chance of being a noise peak.  The Signal and B-mode peaks are very rarely co-incident, but as a conservative measure we define a high confidence sample, (Signal (b)), where the B-mode at the peak must be less than half the amplitude of the signal.}
\label{tab:peakstat}
\end{table}

To define a low-significance $2\sigma < \kappa_{\rm peak} < 4\sigma$ 
substructure sample we use high confidence selection criteria where
the peak must have less 
than 33\% probability of being a random noise fluctuation, and the 
B-mode at the location of the peak must be less than half the amplitude 
of the E-mode.  The last row of Table~\ref{tab:peakstat} lists the number of
peaks that meet these criteria for different significance levels, 
leaving 16 `substructure' peaks with  $2\sigma < \kappa_{\rm peak} < 4\sigma$. 
Note that the 7 peaks with $\kappa_{\rm peak} > 4\sigma$ are all associated 
with the central regions of the 
four main structures in the supercluster, discussed in 
section~\ref{sec:res}, as shown by the marked crosses in 
Figure~\ref{fig:mainstruct} that are enclosed by the $4\sigma$ contour.

Assuming all 16 substructure peaks in the dark matter map are
associated with the supercluster, we can calculate a mass for
these halos using  Equation~\ref{eqn:mass}.  We find an average mass of
$M(<0.75') = 0.35 \pm 0.04 \, \rm{h}^{-1} 10^{13}M_\odot$ for the
$2\sigma < \kappa_{\rm peak} < 3\sigma$ group, and $M(<0.75') = 0.57
\pm 0.06 \, \rm{h}^{-1} 10^{13}M_\odot$ for the $3\sigma < \kappa_{\rm
peak} < 4\sigma$ group.  Table~\ref{tab:sub} lists the number of peaks
that are associated with cluster galaxies where $L(<0.75') > 10^{10}
\rm{h}^{-2} L_{r\odot}$.  We find that over half of the peaks are associated 
with galaxies in the cluster, and provide average mass-to-light ratios 
for these peaks in Table~\ref{tab:sub}.  The mass-to-light ratio of our 
`luminous' substructures are of the same order of
 magnitude as the mass-to-light ratios of the main supercluster structures 
listed in Table~\ref{tab:clusmass}.  

It is likely that many of the peaks 
that are not associated with cluster light are actually at a different 
redshift, as the dark matter map shows the projected surface
mass density along the line of sight.  
We have found one particularly interesting 
$3.5\sigma$ peak in the dark matter distribution to the south west of A901a, 
that is not 
co-incident with any cluster light.  This peak has a 0.1\% chance of 
being a noise fluctuation and is not co-incident with any significant B-modes.  
Our hypothesis is that this peak is due to a mass concentration at a higher redshift than the cluster, supported by the presence of 
a small group of 5 galaxies found within a 0.8 arcmin aperture, 
centred on the $3.5\sigma$ dark matter peak, 
which have photometric redshifts $z = 0.44 \pm 0.04$.  
Intriguingly, out of the four less significant 
$2\sigma - 3\sigma$ dark matter peaks that 
are not associated with cluster light and are 
unlikely to be caused by noise or systematics, we find two peaks that are also 
co-incident with small galaxy groups of 3-4 galaxies at the same redshift $z\sim 0.45$.  
As this is same redshift of the CB1 cluster found in \cite{A9013D}, we are potentially 
seeing extended large-scale structure at higher redshift which is supported by the findings 
of an optical cluster search of COMBO-17 data in this field \citep{Falter}.  This
will be investigated further in a forthcoming three-dimensional analysis.


\begin{table}
\begin{center}
\begin{tabular}{c|c|c|c|c|c}
\hline
$\kappa_{\rm peak}$ & $N_{\rm peak}$ & $\langle M(< 0.75') \rangle$ & $N^{\rm light}_{\rm peak}$ & $\langle M/L \rangle$ \\
& & ($\rm{h}^{-1} 10^{13}M_\odot$) & 
& ($\rm{h} \, M_\odot / L_{r\odot}$)\\\hline
$   2\sigma -   3\sigma $ &  14& $  0.35\pm  0.04$ &   10& $  293\pm   24$ \\
$   3\sigma -   4\sigma $ &   2& $  0.57\pm  0.06$ &    1& $  317\pm  114$ \\

\hline
\end{tabular}
\end{center}
\caption{Mass measurements for the halo substructure sample, 
assuming all peaks in the matter distribution are associated with the 
cluster.  Over half of the peaks have associated cluster galaxies 
($N^{\rm light}_{\rm peak}$), for which we measure a mass-to-light ratio 
($ \langle M/L \rangle$).}
\label{tab:sub}
\end{table}

\section{Discussion and Conclusion}
\label{sec:conc}

From a weak lensing analysis of deep Hubble Space Telescope data, 
we have reconstructed a high-resolution map 
of the dark matter distribution in the Abell 901/902 
supercluster.  We find that the maximal peaks in the dark matter 
distribution are 
very well matched to the locations of the brightest cluster galaxies in the
most massive structures in the supercluster.  These structures are 
A901a, A901b, A902 and the South West group, 
all of which are detected in
our dark matter map at high significance.  

Owing to the high number density of
resolved objects in the {\it HST} data, 
we have been able to produce a map with 
sub-arcminute resolution.  This has allowed us to resolve the 
morphology of the dark matter structures, finding profiles that are far from
the spherically symmetric NFW models that
are typically used to model such systems.  We find local maxima 
in the dark matter distribution around the main structures, that are
also seen in the distribution of galaxies.  Furthermore we see a  
significant extension in the dark matter distribution around A901a, in the 
direction of an in-falling X-ray group called A901$\alpha$ \citep{Gray07}.

We have presented mass, mass-to-light and mass-to-stellar mass ratio estimates
for each of the main structures, finding A901a and A901b to be the most 
massive clusters in the system with 
$M(<1') \sim 2\times10^{13} \rm{h}^{-1} M_\odot$.  
Contrary to the analysis of \citet{meg02} we find
no evidence for the variation of the mass-to-light ratio or the 
mass-to-stellar mass ratio between the different clusters measured in a
1 arcmin aperture ($\sim 120 \rm{h}^{-1}$ kpc) centred on each cluster.  
We have shown the variation of the 
mass-to-stellar mass ratio across the clusters, finding an initial rise in 
$M/M_*$ as a function of distance from the clusters central BCG, followed by a steep 
decrease.

We have investigated the less significant substructures in the dark matter 
map that
are detected at $< 4\sigma$.  Comparing the profile of these peaks with 
what is expected from a random noise peak 
we have selected a sample of substructures where the 
likelihood of those peaks being noise or a result of an 
imperfect PSF correction is low.  
We find that over half of these peaks are associated with galaxies in the 
cluster, yielding mass-to-light ratios that are comparable to the 
mass-to-light ratios found in the main structures in the supercluster.
The remaining peaks in the distribution are likely to be associated with 
galaxy groups at higher redshift \citep{Falter}, 
supported by the discovery of several
co-incident groups of galaxies at $z\sim 0.45$.

One interesting result of \citet{meg02} was a tentative detection of a 
filamentary extension between A901a and A901b.  We do not recover this signal 
in this analysis however and conclude that this 
feature was a result of residual PSF systematics and the KS93 mass 
reconstruction method used in the \citet{meg02} ground-based analysis.  
A similar non-detection and conclusion was 
drawn by \citet{Gavazzi04} on a re-analysis of the tentative lensing
detection of filamentary structure in the MS0302+17 supercluster 
by \citet{K98}.  These two null results do not mean, however, that 
filamentary extensions of dark matter do not exist between clusters. 
Instead, as shown by \citet{Dolag06}, we are finding that 
intra-cluster filaments are 
very difficult to detect through weak lensing.  From numerical simulations, 
\citet{Dolag06} determine an expected filamentary shear signal from a 
supercluster filament of $g \sim 0.01$, 
which is a factor of three smaller than the noise on 1 arcmin scales in this 
{\it HST} analysis.  To detect a signal of this magnitude would require 
significantly deeper space-based observations.  
An alternative, that we are currently 
investigating, is the detection of weak gravitational flexion, 
a third order weak lensing effect 
that will be very effective at probing the
sub-structures that were resolved in this weak shear analysis 
\citep[see for example][]{Bacon_flexion}, and is also a potential 
way to recover more information about intra-cluster filaments.

The dark matter map presented in this paper will form the basis of 
future studies of galaxy morphology and galaxy type 
in an over-dense dark matter 
environment.   Comparing the results
of this analysis with the previous ground-based analysis clearly 
demonstrates the importance of space-based observations for future
high resolution weak lensing dark matter 'observations' of dense environments.

\section{Acknowledgments}
We thank the referee for their useful and detailed comments. 
CH thanks Jasper Wall, Tom Kitching, Tim Schrabback and Martina Kleinheinrich
for useful discussions.  CH
acknowledges the support of a European Commission Programme $6^{\rm
th}$ framework Marie Cure Outgoing International Fellowship under
contract MOIF-CT-2006-21891, and a CITA National fellowship. 
CYP is grateful for support provided 
through STScI and NRC-HIA Fellowship programs. 
MEG was supported by an Anne McLaren Research Fellowship, 
LVW by NSERC, CIfAR and CFI, EFB and KJ by the DFG's Emmy Noether
Programme, AB by the DLR (50 OR 0404), 
MB and EvK by the Austrian Science Foundation FWF under 
grant P18416, SFS by the Spanish MEC grants 
AYA2005-09413-C02-02 and the 
PAI of the Junta de Andaluc\'\i a as research group FQM322,
CW by a PPARC Advanced Fellowship, SJ by NASA under LTSA
Grant NAG5-13063 and NSF under AST-0607748 and DHM by NASA under 
LTSA Grant NAG5-13102. 
Support for STAGES was
provided by NASA through GO-10395 from STScI operated by AURA under
NAS5-26555.

\bibliographystyle{mn2e}
\bibliography{ceh_2007}

\begin{thebibliography}{65}
\expandafter\ifx\csname natexlab\endcsname\relax\def\natexlab#1{#1}\fi

\bibitem[{{Arag{\'o}n-Salamanca et al.}(2008)}]{Aragon}
{Arag{\'o}n-Salamanca et al.}, 2008, In preperation

\bibitem[{Bacon {et~al.}(2006)Bacon, Goldberg, Rowe, \& Taylor}]{Bacon_flexion}
Bacon D., Goldberg D., Rowe B., Taylor A., 2006, MNRAS, 365, 414

\bibitem[{{Balogh} {et~al.}(2000){Balogh}, {Navarro}, \& {Morris}}]{Balogh00}
{Balogh} M.~L., {Navarro} J.~F., {Morris} S.~L., 2000, ApJ, 540, 113

\bibitem[{{Balogh et al.}(2004)}]{Balogh04}
{Balogh et al.} M., 2004, MNRAS, 348, 1355

\bibitem[{{Bartelmann}(1996)}]{Bartelmann96}
{Bartelmann} M., 1996, A\&A, 313, 697

\bibitem[{Bartelmann \& Schneider(2001)}]{Bible}
Bartelmann M., Schneider P., 2001, Physics Reports, 340, 291

\bibitem[{{Baugh} \& {Efstathiou}(1993)}]{BaughEfst}
{Baugh} C.~M., {Efstathiou} G., 1993, MNRAS, 265, 145

\bibitem[{{Bekki}(1999)}]{Bekki}
{Bekki} K., 1999, ApJL, 510, L15

\bibitem[{{Bell} {et~al.}(2007){Bell}, {Zheng}, {Papovich}, {Borch}, {Wolf}, \&
  {Meisenheimer}}]{Bell07}
{Bell} E.~F., {Zheng} X.~Z., {Papovich} C., {Borch} A., {Wolf} C.,
  {Meisenheimer} K., 2007, ApJ, 663, 834

\bibitem[{{Benjamin} {et~al.}(2007){Benjamin}, {Heymans}, {Semboloni}, {van
  Waerbeke}, {Hoekstra}, {Erben}, {Gladders}, {Hetterscheidt}, {Mellier}, \&
  {Yee}}]{JB07}
{Benjamin} J., {Heymans} C., {Semboloni} E., {van Waerbeke} L., {Hoekstra} H.,
  {Erben} T., {Gladders} M.~D., {Hetterscheidt} M., {Mellier} Y., {Yee}
  H.~K.~C., 2007, MNRAS, 381, 702

\bibitem[{{Bertin} \& {Arnouts}(1996)}]{SExt}
{Bertin} E., {Arnouts} S., 1996, A\&AS, 117, 393

\bibitem[{{Blanton} {et~al.}(2006){Blanton}, {Eisenstein}, {Hogg}, \&
  Zehavi}]{Blanton06}
{Blanton} M.~R., {Eisenstein} D., {Hogg} D., Zehavi I., 2006, ApJ, 645, 977

\bibitem[{{Blanton} {et~al.}(2005){Blanton}, {Eisenstein}, {Hogg}, \&
  {Brinkmann}}]{Blanton05}
{Blanton} M.~R., {Eisenstein} D., {Hogg} D.~W.~and{Schlegel} D.~J., {Brinkmann}
  J., 2005, ApJ, 629, 143

\bibitem[{{Bond} \& {Efstathiou}(1987)}]{BE87}
{Bond} J.~R., {Efstathiou} G., 1987, MNRAS, 226, 655

\bibitem[{Borch {et~al.}(2006)Borch, Meisenheimer, Bell, Rix, Wolf, Dye,
  Kleinheinrich, Kovacs, \& Wisotzki}]{Borch}
Borch A., Meisenheimer K., Bell E., Rix H.~W., Wolf C., Dye S., Kleinheinrich
  M., Kovacs Z., Wisotzki L., 2006, A\&A, 453, 869

\bibitem[{{Bullock} {et~al.}(2001){Bullock}, {Kolatt}, {Sigad}, {Somerville},
  {Kravtsov}, {Klypin}, {Primack}, \& {Dekel}}]{Bullock01}
{Bullock} J.~S., {Kolatt} T.~S., {Sigad} Y., {Somerville} R.~S., {Kravtsov}
  A.~V., {Klypin} A.~A., {Primack} J.~R., {Dekel} A., 2001, MNRAS, 321, 559

\bibitem[{{Caldwell} {et~al.}(2008){Caldwell}, {McIntosh}, {Rix}, {Barden},
  {Beckwith}, {Bell}, {Borch}, {Heymans}, {H\"au\ss ler}, {Jahnke}, {Jogee},
  {Meisenheimer}, {Peng}, {Sanchez}, {Somerville}, {Wisotzki}, \&
  {Wolf}}]{CaldwellGEMS}
{Caldwell} J.~A.~R., {McIntosh} D.~H., {Rix} H.-W., {Barden} M., {Beckwith}
  S.~V.~W., {Bell} E.~F., {Borch} A., {Heymans} C., {H\"au\ss ler} B., {Jahnke}
  K., {Jogee} S., {Meisenheimer} K., {Peng} C.~Y., {Sanchez} S.~F.,
  {Somerville} R.~S., {Wisotzki} L., {Wolf} C., 2008, ApJS, 174, 136

\bibitem[{{Chabrier}(2003)}]{Chabrier}
{Chabrier} G., 2003, Publ. Astron. Soc. Pacific, 115, 763

\bibitem[{{Clowe} {et~al.}(2006){Clowe}, {Brada{\v c}}, {Gonzalez},
  {Markevitch}, {Randall}, {Jones}, \& {Zaritsky}}]{Clowe06}
{Clowe} D., {Brada{\v c}} M., {Gonzalez} A.~H., {Markevitch} M., {Randall}
  S.~W., {Jones} C., {Zaritsky} D., 2006, ApJL, 648, L109

\bibitem[{{Clowe} {et~al.}(1998){Clowe}, {Luppino}, {Kaiser}, {Henry}, \&
  {Gioia}}]{Clowe98}
{Clowe} D., {Luppino} G.~A., {Kaiser} N., {Henry} J.~P., {Gioia} I.~M., 1998,
  ApJL, 497, L61+

\bibitem[{Crittenden {et~al.}(2002)Crittenden, Natarajan, Pen, \&
  Theuns}]{CNPT02}
Crittenden R., Natarajan R., Pen U., Theuns T., 2002, ApJ, 568, 20

\bibitem[{{Dietrich} {et~al.}(2005){Dietrich}, {Schneider}, {Clowe},
  {Romano-D{\'{\i}}az}, \& {Kerp}}]{Dietrich}
{Dietrich} J.~P., {Schneider} P., {Clowe} D., {Romano-D{\'{\i}}az} E., {Kerp}
  J., 2005, A\&A, 440, 453

\bibitem[{{Dolag} {et~al.}(2004){Dolag}, {Bartelmann}, {Perrotta},
  {Baccigalupi}, {Moscardini}, {Meneghetti}, \& {Tormen}}]{Dolag04}
{Dolag} K., {Bartelmann} M., {Perrotta} F., {Baccigalupi} C., {Moscardini} L.,
  {Meneghetti} M., {Tormen} G., 2004, A\&A, 416, 853

\bibitem[{{Dolag} {et~al.}(2006){Dolag}, {Meneghetti}, {Moscardini}, {Rasia},
  \& {Bonaldi}}]{Dolag06}
{Dolag} K., {Meneghetti} M., {Moscardini} L., {Rasia} E., {Bonaldi} A., 2006,
  MNRAS, 370, 656

\bibitem[{{Dressler}(1980)}]{Dressler}
{Dressler} A., 1980, ApJ, 236, 351

\bibitem[{{Eke} {et~al.}(2001){Eke}, {Navarro}, \& {Steinmetz}}]{Eke01}
{Eke} V.~R., {Navarro} J.~F., {Steinmetz} M., 2001, ApJ, 554, 114

\bibitem[{Falter {et~al.}(2008)Falter, Roeser, {Wolf}, \& Hippelein}]{Falter}
Falter S., Roeser H., {Wolf} C., Hippelein H., 2008, {In preperation}

\bibitem[{{Gavazzi} {et~al.}(2004){Gavazzi}, {Mellier}, {Fort}, {Cuillandre},
  \& {Dantel-Fort}}]{Gavazzi04}
{Gavazzi} R., {Mellier} Y., {Fort} B., {Cuillandre} J.-C., {Dantel-Fort} M.,
  2004, A\&A, 422, 407

\bibitem[{{Gilmour} {et~al.}(2007){Gilmour}, {Gray}, {Almaini}, {Best}, {Wolf},
  {Meisenheimer}, {Papovich}, \& {Bell}}]{Gilmour}
{Gilmour} R., {Gray} M.~E., {Almaini} O., {Best} P., {Wolf} C., {Meisenheimer}
  K., {Papovich} C., {Bell} E., 2007, MNRAS, 380, 1467

\bibitem[{{Gorenstein} {et~al.}(1988){Gorenstein}, {Shapiro}, \&
  {Falco}}]{masssheet}
{Gorenstein} M.~V., {Shapiro} I.~I., {Falco} E.~E., 1988, ApJ, 327, 693

\bibitem[{Gray {et~al.}(2008)Gray, Gilmour, \& collaborators}]{Gray07}
Gray M., Gilmour R., collaborators, 2008, In preperation

\bibitem[{Gray {et~al.}(2002)Gray, Taylor, Meisenheimer, Dye, Wolf, \&
  Thommes}]{meg02}
Gray M., Taylor A., Meisenheimer K., Dye S., Wolf C., Thommes E., 2002, ApJ,
  568, 141

\bibitem[{{Gray} {et~al.}(2004){Gray}, {Wolf}, {Meisenheimer}, {Taylor}, {Dye},
  {Borch}, \& {Kleinheinrich}}]{Gray04}
{Gray} M.~E., {Wolf} C., {Meisenheimer} K., {Taylor} A., {Dye} S., {Borch} A.,
  {Kleinheinrich} M., 2004, MNRAS, 347, L73

\bibitem[{{Gray et al.}(2008)}]{GrayPeng}
{Gray et al.} M., 2008, In preperation

\bibitem[{{Gunn} \& {Gott}(1972)}]{GG72}
{Gunn} J.~E., {Gott} J.~R.~I., 1972, ApJ, 176, 1

\bibitem[{Heymans {et~al.}(2006)Heymans, {Bell}, {Rix}, {Barden}, {Borch},
  {Caldwell}, {McIntosh}, {Meisenheimer}, {Peng}, {Wolf}, {Beckwith}, ,
  {H\"au\ss ler}, {Jahnke}, {Jogee}, , {Sanchez}, {Somerville}, \&
  {Wisotzki}}]{GEMSgg}
Heymans C., {Bell} E.~F., {Rix} H.-W., {Barden} M., {Borch} A., {Caldwell}
  J.~A.~R., {McIntosh} D.~H., {Meisenheimer} K., {Peng} C.~Y., {Wolf} C.,
  {Beckwith} S.~V.~W., , {H\"au\ss ler} B., {Jahnke} K., {Jogee} S., ,
  {Sanchez} S.~F., {Somerville} R.~S., {Wisotzki} L., 2006, MNRAS, 371L, 60

\bibitem[{Heymans {et~al.}(2005)Heymans, Brown, {Barden}, {Caldwell}, {Jahnke},
  {Rix}, Taylor, {Beckwith}, {Bell}, {Borch}, {H\"au\ss ler}, {Jogee},
  {McIntosh}, {Meisenheimer}, {Peng}, {Sanchez}, {Somerville}, {Wisotzki}, \&
  {Wolf}}]{HymzGEMS}
Heymans C., Brown M.~L., {Barden} M., {Caldwell} J.~A.~R., {Jahnke} K., {Rix}
  H.-W., Taylor A.~N., {Beckwith} S.~V.~W., {Bell} E.~F., {Borch} A., {H\"au\ss
  ler} B., {Jogee} S., {McIntosh} D.~H., {Meisenheimer} K., {Peng} C.~Y.,
  {Sanchez} S.~F., {Somerville} R.~S., {Wisotzki} L., {Wolf} C., 2005, MNRAS,
  160

\bibitem[{{Heymans} {et~al.}(2006){Heymans}, {Van Waerbeke}, {Bacon}, {Berge},
  {Bernstein}, {Bertin}, {Bridle}, {Brown}, {Clowe}, {Dahle}, {Erben}, {Gray},
  {Hetterscheidt}, {Hoekstra}, {Hudelot}, {Jarvis}, {Kuijken}, {Margoniner},
  {Massey}, {Mellier}, {Nakajima}, {Refregier}, {Rhodes}, {Schrabback}, \&
  {Wittman}}]{HymzSTEP}
{Heymans} C., {Van Waerbeke} L., {Bacon} D., {Berge} J., {Bernstein} G.,
  {Bertin} E., {Bridle} S., {Brown} M.~L., {Clowe} D., {Dahle} H., {Erben} T.,
  {Gray} M., {Hetterscheidt} M., {Hoekstra} H., {Hudelot} P., {Jarvis} M.,
  {Kuijken} K., {Margoniner} V., {Massey} R., {Mellier} Y., {Nakajima} R.,
  {Refregier} A., {Rhodes} J., {Schrabback} T., {Wittman} D., 2006, MNRAS, 368,
  1323

\bibitem[{{Hoekstra}(2007)}]{HH07}
{Hoekstra} H., 2007, MNRAS, 379, 317

\bibitem[{{Hoekstra} {et~al.}(2005){Hoekstra}, {Hsieh}, {Yee}, {Lin}, \&
  {Gladders}}]{HHbf05}
{Hoekstra} H., {Hsieh} B.~C., {Yee} H.~K.~C., {Lin} H., {Gladders} M.~D., 2005,
  ApJ, 635, 73

\bibitem[{{Kaiser} \& {Squires}(1993)}]{KS93}
{Kaiser} N., {Squires} G., 1993, ApJ, 404, 441

\bibitem[{Kaiser {et~al.}(1995)Kaiser, Squires, \& Broadhurst}]{KSB}
Kaiser N., Squires G., Broadhurst T., 1995, ApJ, 449, 460

\bibitem[{{Kaiser} {et~al.}(1998){Kaiser}, {Wilson}, {Luppino}, {Kofman},
  {Gioia}, {Metzger}, \& {Dahle}}]{K98}
{Kaiser} N., {Wilson} G., {Luppino} G., {Kofman} L., {Gioia} I., {Metzger} M.,
  {Dahle} H., 1998, ArXiv Astrophysics e-prints astro-ph 9809268

\bibitem[{{Kleinheinrich} {et~al.}(2006){Kleinheinrich}, {Schneider}, {Rix},
  {Erben}, {Wolf}, {Schirmer}, {Meisenheimer}, {Borch}, {Dye}, {Kovacs}, \&
  {Wisotzki}}]{MK06}
{Kleinheinrich} M., {Schneider} P., {Rix} H.~W., {Erben} T., {Wolf} C.,
  {Schirmer} M., {Meisenheimer} K., {Borch} A., {Dye} S., {Kovacs} Z.,
  {Wisotzki} L., 2006, A\&A, 445, 441

\bibitem[{{Kroupa}(2001)}]{Kroupa01}
{Kroupa} P., 2001, MNRAS, 322, 231

\bibitem[{{Kroupa} {et~al.}(1993){Kroupa}, {Tout}, \& {Gilmore}}]{Kroupa}
{Kroupa} P., {Tout} C.~A., {Gilmore} G., 1993, MNRAS, 262, 545

\bibitem[{{Lane} {et~al.}(2007){Lane}, {Gray}, {Arag{\'o}n-Salamanca}, {Wolf},
  \& {Meisenheimer}}]{Lane}
{Lane} K.~P., {Gray} M.~E., {Arag{\'o}n-Salamanca} A., {Wolf} C.,
  {Meisenheimer} K., 2007, MNRAS, 378, 716

\bibitem[{{Larson} {et~al.}(1980){Larson}, {Tinsley}, \& {Caldwell}}]{Larson}
{Larson} R.~B., {Tinsley} B.~M., {Caldwell} C.~N., 1980, ApJ, 237, 692

\bibitem[{{Luppino} \& {Kaiser}(1997)}]{LK97}
{Luppino} G.~A., {Kaiser} N., 1997, ApJ, 475, 20

\bibitem[{{Mahdavi} {et~al.}(2007){Mahdavi}, {Hoekstra}, {Babul}, {Balam}, \&
  {Capak}}]{Mahdavi}
{Mahdavi} A., {Hoekstra} H., {Babul} A., {Balam} D.~D., {Capak} P.~L., 2007,
  ApJ, 668, 806

\bibitem[{{Mandelbaum} {et~al.}(2006){Mandelbaum}, {Seljak}, {Kauffmann},
  {Hirata}, \& {Brinkmann}}]{RM06}
{Mandelbaum} R., {Seljak} U., {Kauffmann} G., {Hirata} C.~M., {Brinkmann} J.,
  2006, MNRAS, 368, 715

\bibitem[{{Massey} {et~al.}(2007){Massey}, {Heymans}, {Berg{\'e}}, {Bernstein},
  {Bridle}, {Clowe}, {Dahle}, {Ellis}, {Erben}, {Hetterscheidt}, {High},
  {Hirata}, {Hoekstra}, {Hudelot}, {Jarvis}, {Johnston}, {Kuijken},
  {Margoniner}, {Mandelbaum}, {Mellier}, {Nakajima}, {Paulin-Henriksson},
  {Peeples}, {Roat}, {Refregier}, {Rhodes}, {Schrabback}, {Schirmer}, {Seljak},
  {Semboloni}, \& {Van Waerbeke}}]{STEP2}
{Massey} R., {Heymans} C., {Berg{\'e}} J., {Bernstein} G., {Bridle} S., {Clowe}
  D., {Dahle} H., {Ellis} R., {Erben} T., {Hetterscheidt} M., {High} F.~W.,
  {Hirata} C., {Hoekstra} H., {Hudelot} P., {Jarvis} M., {Johnston} D.,
  {Kuijken} K., {Margoniner} V., {Mandelbaum} R., {Mellier} Y., {Nakajima} R.,
  {Paulin-Henriksson} S., {Peeples} M., {Roat} C., {Refregier} A., {Rhodes} J.,
  {Schrabback} T., {Schirmer} M., {Seljak} U., {Semboloni} E., {Van Waerbeke}
  L., 2007, MNRAS, 376, 13

\bibitem[{{Moore} {et~al.}(1996){Moore}, {Katz}, {Lake}, {Dressler}, \&
  {Oemler}}]{Moore96}
{Moore} B., {Katz} N., {Lake} G., {Dressler} A., {Oemler} A., 1996, Nature,
  379, 613

\bibitem[{{Moore} {et~al.}(1998){Moore}, {Lake}, \& {Katz}}]{Moore98}
{Moore} B., {Lake} G., {Katz} N., 1998, ApJ, 495, 139

\bibitem[{{Navarro} {et~al.}(1997){Navarro}, {Frenk}, \& {White}}]{NFW97}
{Navarro} J.~F., {Frenk} C.~S., {White} S.~D.~M., 1997, ApJ, 490, 493

\bibitem[{{Rhodes} {et~al.}(2007){Rhodes}, {Massey}, {Albert}, {Collins},
  {Ellis}, {Heymans}, {Gardner}, {Kneib}, {Koekemoer}, {Leauthaud}, {Mellier},
  {Refregier}, {Taylor}, \& {Van Waerbeke}}]{RhodesPSF}
{Rhodes} J.~D., {Massey} R., {Albert} J., {Collins} N., {Ellis} R.~S.,
  {Heymans} C., {Gardner} J.~P., {Kneib} J.-P., {Koekemoer} A., {Leauthaud} A.,
  {Mellier} Y., {Refregier} A., {Taylor} J.~E., {Van Waerbeke} L., 2007, ApJS,
  172, 203

\bibitem[{{Rix} {et~al.}(2004){Rix}, {Barden}, {Beckwith}, {Bell}, {Borch},
  {Caldwell}, {H\"au\ss ler}, {Jahnke}, {Jogee}, {McIntosh}, {Meisenheimer},
  {Peng}, {Sanchez}, {Somerville}, {Wisotzki}, \& {Wolf}}]{RixGEMS}
{Rix} H.-W., {Barden} M., {Beckwith} S.~V.~W., {Bell} E.~F., {Borch} A.,
  {Caldwell} J.~A.~R., {H\"au\ss ler} B., {Jahnke} K., {Jogee} S., {McIntosh}
  D.~H., {Meisenheimer} K., {Peng} C.~Y., {Sanchez} S.~F., {Somerville} R.~S.,
  {Wisotzki} L., {Wolf} C., 2004, ApJS, 152, 163

\bibitem[{{Schneider} \& {Rix}(1997)}]{SchRix}
{Schneider} P., {Rix} H., 1997, ApJ, 474, 25

\bibitem[{{Schrabback} {et~al.}(2007){Schrabback}, {Erben}, {Simon},
  {Miralles}, {Schneider}, {Heymans}, {Eifler}, {Fosbury}, {Freudling},
  {Hetterscheidt}, {Hildebrandt}, \& {Pirzkal}}]{SchPSF}
{Schrabback} T., {Erben} T., {Simon} P., {Miralles} J.-M., {Schneider} P.,
  {Heymans} C., {Eifler} T., {Fosbury} R.~A.~E., {Freudling} W.,
  {Hetterscheidt} M., {Hildebrandt} H., {Pirzkal} N., 2007, A\&A, 468

\bibitem[{{Taylor} {et~al.}(2004){Taylor}, {Bacon}, {Gray}, {Wolf},
  {Meisenheimer}, {Dye}, {Borch}, {Kleinheinrich}, {Kovacs}, \&
  {Wisotzki}}]{A9013D}
{Taylor} A.~N., {Bacon} D.~J., {Gray} M.~E., {Wolf} C., {Meisenheimer} K.,
  {Dye} S., {Borch} A., {Kleinheinrich} M., {Kovacs} Z., {Wisotzki} L., 2004,
  MNRAS, 353, 1176

\bibitem[{{Van Waerbeke}(2000)}]{VW2000}
{Van Waerbeke} L., 2000, MNRAS, 313, 524

\bibitem[{{Wolf} {et~al.}(2005){Wolf}, {Gray}, \& {Meisenheimer}}]{wgm}
{Wolf} C., {Gray} M.~E., {Meisenheimer} K., 2005, A\&A, 443, 435

\bibitem[{Wolf {et~al.}(2004)Wolf, Meisenheimer, Kleinheinrich, Borch, Dye,
  Gray, Wisotzki, Bell, Rix, Hasinger, \& Szokoly}]{C17}
Wolf C., Meisenheimer K., Kleinheinrich M., Borch A., Dye S., Gray M., Wisotzki
  L., Bell E., Rix H.-W., Hasinger A. C.~G., Szokoly G., 2004, A\&A, 421, 913

\bibitem[{Wolf {et~al.}(2003)Wolf, Meisenheimer, Rix, Borch, Dye, \&
  Kleinheinrich}]{Wolf03}
Wolf C., Meisenheimer K., Rix H.-W., Borch A., Dye S., Kleinheinrich M., 2003,
  A\&A, 401, 73

\bibitem[{{Wright} \& {Brainerd}(2000)}]{WB00}
{Wright} C.~O., {Brainerd} T.~G., 2000, ApJ, 534, 34

\end{thebibliography}
\label{lastpage}

\end{document}